\documentclass[11pt]{article}

   \usepackage{graphicx}
   \usepackage{amsfonts}
   \usepackage{amssymb}
   \usepackage{amsthm}
   \usepackage{amsmath}

   \usepackage[active]{srcltx}

   \usepackage{here}
   \usepackage{cite}
   \usepackage{color}
   \usepackage{soul}

   \usepackage[affil-it]{authblk}

   \graphicspath{{figs/}}

\newcommand{\be}{\begin{equation}}
\newcommand{\ee}{\end{equation}}

\newcommand{\beq}{\begin{equation}}
\newcommand{\eeq}{\end{equation}}
\newcommand{\non}{\nonumber}
\newcommand\bea{\begin{eqnarray}}
\newcommand\eea{\end{eqnarray}}

\allowdisplaybreaks
\newcommand\al{\alpha}

\newcommand\tb{\tan\beta}

\newcommand\ReDiag{\mathop{%
  \raise .5pt\hbox{[}%
  \widetilde{\mathrm{Re}}%
  \raise .5pt\hbox{]}}}
\newcommand\ReOffDiag{\mathop{%
  \raise .5pt\hbox{$\llbracket$}%
  \widetilde{\mathrm{Re}}%
  \raise .5pt\hbox{$\rrbracket$}}}

\newcommand\Mh{M_h}
\newcommand\MH{M_H}
\newcommand\MA{M_A}
\newcommand\MHp{M_{H^\pm}}

\newcommand\mb{m_b}
\newcommand\mt{m_t}

\newcommand\gl{{\tilde g}}
\newcommand\mgl{m_\gl}

\newcommand\ino[1]{\tilde\chi_{#1}}

\newcommand\chapm[1]{\ino{#1}^\pm}

\newcommand\mcha[1]{m_{\chapm{#1}}}

\newcommand\neu[1]{\ino{#1}^0}
\newcommand\mneu[1]{m_{\neu{#1}}}

\newcommand\refeq[1]{Eq.~(\ref{#1})}

\newcommand\refta[1]{Tab.~\ref{#1}}
\newcommand\refse[1]{Sect.~\ref{#1}}

\newcommand\citere[1]{Ref.~\cite{#1}}
\newcommand\citeres[1]{Refs.~\cite{#1}}

\newcommand{\CP}{{\cal CP}}
\newcommand{\cp}{{\CP}}

\newcommand{\tev}{\,\, \mathrm{TeV}}
\newcommand{\gev}{\,\, \mathrm{GeV}}
\newcommand{\mev}{\,\, \mathrm{MeV}}

\newcommand\MO{\texttt{MicrOMEGAs}}

\newcommand{\br}{\text{BR}}

\newcommand{\sig}{\sigma}

\def\order#1{\ensuremath{{\cal O}(#1)}}
\def\reffi#1{\mbox{Fig.~\ref{#1}}}

\def\ga{\gamma}

\newcommand{\VL}{\left( \begin{array}{c}}
\newcommand{\VR}{\end{array} \right)}
\newcommand{\ML}{\left( \begin{array}{cc}}
\newcommand{\MLd}{\left( \begin{array}{ccc}}
\newcommand{\MLv}{\left( \begin{array}{cccc}}
\newcommand{\MR}{\end{array} \right)}

\definecolor{Lightblue}{cmyk}{0.9,0.1,0.1,0.3}
\definecolor{dgelborange}{cmyk}{0.,0.3,0.5, 0.}
\definecolor{Orange}{cmyk}{0.,0.5,0.5, 0.}
\definecolor{Lila}{rgb}{0.5,0.,1}

\newcommand{\complex}{{{\rm I} \kern -.59em {\rm C}}}

\newcommand{\nn}{\nonumber}

\begin{document}

\thispagestyle{empty}

\title{
\vspace*{-3.5cm}
Updates and New Results in Models with Reduced Couplings}
\date{}
\author{ \hspace*{7mm} S. Heinemeyer$^{1,2,3}$\thanks{email: Sven.Heinemeyer@cern.ch} , M. Mondrag\'on$^4$\thanks{email: myriam@fisica.unam.mx} ,
G. Patellis$^5$\thanks{email: patellis@central.ntua.gr} ,
N. Tracas$^5$\thanks{email: ntrac@central.ntua.gr}$\,$  and G. Zoupanos$^{5,6,7,8}$\thanks{email: George.Zoupanos@cern.ch}\\
{\small
$^{1}$  Instituto de F\'{\i}sica Te\'{o}rica (UAM/CSIC), Universidad Aut\'{o}noma de Madrid Cantoblanco, 28049 Madrid, Spain\\
$^{2}$  Campus of International Excellence UAM+CSIC, Cantoblanco, 28049 Madrid, Spain\\
$^3$ Instituto de F\'{\i}sica de Cantabria (CSIC-UC), E-39005 Santander, Spain \\
$^4$  Instituto de F\'{\i}sica, Universidad Nacional Aut\'onoma de M\'exico, A.P. 20-364, CDMX 01000 M\'exico \\
$^5$  Physics Department, National Technical University, 157 80 Zografou, Greece\\
$^6$  Max-Planck Institut f\"ur Physik, F\"ohringer Ring 6, D-80805 M\"unchen, Germany\\
$^7$  Theoretical Physics Department, CERN, Geneva, Switzerland\\
$^8$  Max-Planck-Institut f\"ur Gravitationsphysik (Albert-Einstein-Institut), M\"uhlenberg 1, D-14476 Potsdam, Germany  
}
}

\maketitle


\abstract{The idea of reduction of couplings consists in searching for
  renormalization  group invariant  relations between parameters of a
  renormalizable theory that hold to all orders of perturbation
  theory. Based on the principle of the reduction of couplings, one can
  construct Finite Unified Theories (FUTs) which are
  $N=1$ supersymmetric Grand Unified Theories that can
  be made all-order finite. The prediction of the top quark mass well in
  advance of its experimental discovery and the prediction of the light
  Higgs boson mass in the range $\sim 121-126$ GeV
  much earlier than its experimental discovery are among the celebrated
  successes of such models. Here, after a brief review of the reduction
  of couplings method and the properties of the resulting finiteness in
  supersymmetric theories, we analyse four phenomenologically favoured
  models: a minimal version of the $N=1$
  $SU(5)$, a finite $N=1$
  $SU(5)$, a $N=1$ finite
  $SU(3)\otimes SU(3)\otimes SU(3)$ model and a
  reduced version of the Minimal Supersymmetric Standard Model
  (MSSM). A relevant update in the phenomenological evaluation has been
  the improved light Higgs-boson mass prediction as provided by the
  latest version of {\tt FeynHiggs}. 
  All four models predict relatively heavy supersymmetric
  spectra that start just below or above the TeV~scale, consistent with
  the non-observation LHC results. Depending on the model, the 
  lighter regions of the spectra could be accessible at CLIC, while the
  FCC-hh will be able to 
  test large parts of predicted spectrum of each model. The lightest
  supersymmetric particle (LSP), which is a neutralino, is considered as
  a cold dark matter candidate and put to test using the latest
  \MO~code. }

\vspace{0.5cm}

\noindent
{\tt 
IFT-UAM/CSIC-20-026
}

\vfill
\newpage

\section{Introduction}
During the last years, a series of successes in developing frameworks such as String Theories and
Noncommutativity have been presented, as a result of various  theoretical endeavours that aim to describe~the fundamental~theory at the Planck~scale. 
However, the essence of all theoretical efforts in Particle Physics is
to understand the free parameters of the Standard Model (SM) in terms of fewer, fundamental ones. In other words, to achieve a {\it reduction of couplings} \cite{book}. Unfortunately, the several recent successes in the above frameworks do not offer anything concerning the understanding of the SM free parameters. The problem of the large number of free~parameters is deeply
connected to the infinities that are present at the quantum level.
The renormalization programme removes infinities by introducing
counterterms, but it does so at the big cost of leaving the~corresponding terms as~free parameters.

Although the SM is successful in describing
elementary particles and their interactions, it is a widespread belief that it should be the low-energy limit of a fundamental theory. The search for beyond the Standard Model (BSM) theories expands in various~directions. One of the most efficient ways to reduce the plethora of
free parameters of a theory (and thus render it more predictive) is the introduction of a symmetry. A very good example of such a procedure are the Grand Unified Theories (GUTs) \cite{Pati:1973rp,Georgi:1974sy,Georgi:1974yf,Fritzsch:1974nn,Gursey:1975ki,Achiman:1978vg}.
In the  early days, the minimal $SU(5)$ (because of an -approximate- gauge unification) reduced the gauge couplings of the SM, predicting one of them. 
It was the addition of an $N=1$ global (softly broken) supersymmetry (SUSY)
\cite{Amaldi:1991cn,Dimopoulos:1981zb,Sakai:1981gr} that made the prediction viable. In the framework of GUTs, the Yukawa couplings can also be related among themselves. Again, $SU(5)$~demonstrated this by predicting the ratio of the tau to bottom mass \cite{Buras:1977yy} for the SM.
Unfortunately, the requirement of additional gauge symmetry does not seem to help, since new 
complications due to new degrees of freedom arise. 

One could relate seemingly independent parameters via the reduction of couplings method \cite{Zimmermann:1984sx,Oehme:1984yy,Oehme:1985jy} (
see also \cite{Ma:1977hf,Chang:1974bv,Nandi:1978fw}).
This technique reduces the number of free couplings by relating all or some of the couplings of the theory to a single parameter, the ``primary coupling''. This method can identify previously hidden symmetries in a system, but it can also be applied in models with no apparent~symmetry. It is necessary, though to make two assumptions: first, both the
original and the reduced theory have to be renormalizable and second, the relations among parameters should be renormalization group invariant (RGI).

A natural continuation of the idea of Grand Unification is to achieve gauge-Yukawa Unification (GYU), that is to relate the gauge sector to the Yukawa sector. This is a feature of theories in which  \textit{reduction of couplings} has been~applied. The original suggestion for the reduction of couplings in GUTs leads to the search for RGI relations that hold below the Planck scale, which  are in turn preserved down to the
unification scale. Impressively, this observation guarantees 
validity of such relations to all-orders in perturbation theory. This is done by studying
their uniqueness  at one-loop level.
Even more remarkably,
one can find such RGI relations that result in all-loop finiteness. In the sections to follow we will show cases in which these principles will be applied in $N = 1$ SUSY GUTs. The application of the GYU programme in the dimensionless couplings of such theories has been very successful, including the prediction of the mass of the top quark in the minimal
\cite{Kubo:1994bj} (see also Section \ref{se:minimal} for the latest update) and in the finite $N = 1$ $SU(5)$ \cite{Kapetanakis:1992vx,Mondragon:1993tw} before~its experimental discovery \cite{Lancaster:2011wr}.

In order to successfully apply the above-mentioned programme, SUSY appears to be essential. However, one has to understand its breaking as well. This naturally leads to the extention of this search for RGI expressions to the SUSY breaking (SSB) sector of these models, which involves
couplings of dimension one and two. There has been crucial progress on the renormalization properties of the SSB parameters based on the powerful~supergraph method for~studying SUSY theories, applied to the softly broken~ones using the
spurion external~space-time independent~superfields.
In this method a softly broken supersymmetric theory is taken to be 
supersymmetric, where the various couplings and masses are considered external superfields. 
Then, relations among the soft term renormalization
and that of an unbroken SUSY have been derived. 

The application of \textit{reduction of couplings} on $N=1$ SUSY theories has led to many interesting phenomenological developments. In past work, the assumption of introducing a ``universal'' set of soft scalar masses that serve as one of the constraints preserving  two-loop finiteness exhibited a number of problems due to its restrictive~nature. Subsequently, this constraint was replaced by a more ``relaxed'' all-loop RGI sum rule that keeps the most attractive features of the universal case and overcomes the unpleasant~phenomenological consequences. This arsenal of tools and results opened the way for the study of full finite models with few free parameters, with  emphasis given on predictions for the SUSY spectrum and the light Higgs mass.

The Higgs mass prediction, that coincided with the LHC results (ATLAS \cite{Aad:2012tfa,ATLAS:2013mma}
and CMS \cite{Chatrchyan:2012ufa,Chatrchyan:2013lba}) - combined with a predicted relatively heavy spectrum that is consistent with the as of yet non-observation of
SUSY particles - was a success of the all-loop finite $N = 1$ SUSY $SU(5)$ model \cite{Heinemeyer:2007tz}. This case will be
presented in \refse{se:futb}, while the results for another finite theory, namely the $N=1$ (two-loop) finite $SU(3)\otimes SU(3)\otimes SU(3)$, will be presented in \refse{se:trinification}. Furthermore, the above programme was also
applied in the MSSM, with successful results concerning the top, bottom and Higgs masses, also featuring a relatively heavy SUSY spectrum that accommodates a dark matter candidate as well. These results will be presented in \refse{se:rmssm}. The calculation of the lightest Higgs boson mass is done with the (new) {\tt FeynHiggs} code~\cite{Degrassi:2002fi,BHHW,FeynHiggs,Bahl:2019hmm}.

Last but not least, it~is a~well known fact that the lightest~neutralino, being the Lightest
SUSY Particle (LSP), is an
excellent~candidate for Cold Dark~Matter (CDM)~\cite{EHNOS}. Our
analyses presented in \refse{se:minimal}, \ref{se:futb},
\ref{se:trinification} and \ref{se:rmssm}  also include the calculation
of the CDM relic density for each model, using the \MO~code
\cite{Belanger:2001fz,Belanger:2004yn,Barducci:2016pcb}. As will be
discussed, none of the models satisfy the experimental bounds of
the relic density exactly.


\section{Theoretical Basis}\label{se:theory}

\subsection{Reduction of Dimensionless Parameters}
We start by reviewing the basic \textit{reduction of couplings} idea. The aim is, if possible, to express the parameters of a theory that are considered as free in terms of one basic parameter, which we call primary. The basic idea is to search for RGI relations among couplings and use them to reduce the seemingly independent parameters.
Any RGI relation among parameters $g_1,\cdots,g_A$ of a given renormalizable theory can be expressed implicitly as $\Phi (g_1,\cdots,g_A) ~=~\mbox{const.}$
This expression must satisfy the  partial~differential equation~(PDE)
\beq
\mu\,\frac{d \Phi}{d \mu} = {\vec \nabla}\Phi\cdot {\vec \beta} ~=~
\sum_{a=1}^{A}
\,\beta_{a}\,\frac{\partial \Phi}{\partial g_{a}}~=~0~,
\eeq
with $\beta_a$ the  $\beta$-functions of $g_a$.
Solving this PDE is equivalent
to solving a set of ordinary differential equations (ODEs),
the reduction~equations (REs) \cite{Zimmermann:1984sx,Oehme:1984yy,Oehme:1985jy},
\beq
\beta_{g} \,\frac{d g_{a}}{d g} =\beta_{a}~,~a=1,\cdots,A-1~,
\label{redeq}
\eeq
Here, $g$ and  $\beta_{g}$ are the primary
coupling~and~its $\beta$-function, respectively.
The  $\Phi_a$'s can impose up to ($A-1$) independent
RGI~constraints in the  $A$-dimensional parameter~space. As a result, all couplings can be (in principle) expressed in terms of the primary coupling $g$.

This is not enough, as  the number of integration constants of the general solutions of \refeq{redeq} matches the~number of these equations, meaning that we just~traded~an integration constant~for each ordinary~renormalized~coupling, and therefore these cannot be considered as reduced solutions.

The~crucial requirement is the demand that the~REs admit power~series~solutions,
\beq
g_{a} = \sum_{n}\rho_{a}^{(n)}\,g^{2n+1}~,
\label{powerser}
\eeq
that~preserve perturbative~renormalizability.
This way, the integration constant corresponding to each RE is fixed and the RE is
picked up as a special solution out of the set of the general ones.
It is worth noting that a one-loop level examination is enough to decide for the uniqueness of these solutions \cite{Zimmermann:1984sx,Oehme:1984yy,Oehme:1985jy}.  
As an illustration~ on the above, we assume  $\beta$-functions of the~form
\beq
\begin{split}
\beta_{a} &=\frac{1}{16 \pi^2}\left[ \sum_{b,c,d\neq
  g}\beta^{(1)\,bcd}_{a}g_b g_c g_d+
\sum_{b\neq g}\beta^{(1)\,b}_{a}g_b g^2\right]+\cdots~,\\
\beta_{g} &=\frac{1}{16 \pi^2}\beta^{(1)}_{g}g^3+ \cdots~,
\end{split}
\eeq
Here $\cdots$ stands for higher order terms~and  $ \beta^{(1)\,bcd}_{a}$'s
are~symmetric~in $ b,c$ and $d$. We assume~that
$\rho_{a}^{(n)}$ with $n\leq r$ are already determined~uniquely. 
In order to~obtain
$\rho_{a}^{(r+1)}$, the  power series (\ref{powerser}) are inserted into
the~REs (\ref{redeq}) and we collect~terms of ${\cal O}(g^{2r+3})$. Thus, we find
\beq
\sum_{d\neq g}M(r)_{a}^{d}\,\rho_{d}^{(r+1)} = \mbox{lower
  order quantities}~,\non
\eeq
where the  right-hand side is~known by~assumption and
\begin{align}
M(r)_{a}^{d} &=3\sum_{b,c\neq
  g}\,\beta^{(1)\,bcd}_{a}\,\rho_{b}^{(1)}\,
\rho_{c}^{(1)}+\beta^{(1)\,d}_{a}
-(2r+1)\,\beta^{(1)}_{g}\,\delta_{a}^{d}~,\label{M}\\
0 &=\sum_{b,c,d\neq g}\,\beta^{(1)\,bcd}_{a}\,
\rho_{b}^{(1)}\,\rho_{c}^{(1)}\,\rho_{d}^{(1)} +\sum_{d\neq
  g}\beta^{(1)\,d}_{a}\,\rho_{d}^{(1)}
-\beta^{(1)}_{g}\,\rho_{a}^{(1)}~.
\end{align}
Therefore,~the  $\rho_{a}^{(n)}$ for~all $n > 1$ for~a
given set of $\rho_{a}^{(1)}$ can be~uniquely~determined if $\det
M(n)_{a}^{d} \neq 0$ for~all $n \geq 0$. This is checked in all models that reductions of couplings is applied.

The search for power~series solutions to the REs like (\ref{powerser}) is more than justified in SUSY theories, where parameters often behave asymptotically in a similar way.
This ``completely reduced'' theory features only one~independent parameter, rendering this unification very attractive. It is often unrealistic, however, and, usually, fewer RGI constraints are imposed, leading to a partial~reduction \cite{Kubo:1985up,Kubo:1988zu}.

All the above give rise to hints towards an underlying connection among~the requirement~of reduction~of couplings~and
SUSY.  

As an example, we consider a $SU(N)$ gauge theory with $\phi^{i}({\bf N})$ and $\hat{\phi}_{i}(\overline{\bf N})$ complex scalars,
$\psi^{i}({\bf N})$ and $\hat{\psi}_{i}(\overline{\bf N})$ left-handed Weyl spinors
and $\lambda^a (a=1,\dots,N^2-1)$ 
right-handed Weyl spinors~in the adjoint~representation of $SU(N)$.

The Lagrangian (kinetic terms are omitted) includes
\beq
{\cal L} \supset 
i \sqrt{2} \{~g_Y\overline{\psi}\lambda^a T^a \phi
-\hat{g}_Y\overline{\hat{\psi}}\lambda^a T^a \hat{\phi}
+\mbox{h.c.}~\}-V(\phi,\overline{\phi}),
\eeq
where
\beq
V(\phi,\overline{\phi}) =
\frac{1}{4}\lambda_1(\phi^i \phi^{*}_{i})^2+
\frac{1}{4}\lambda_2(\hat{\phi}_i \hat{\phi}^{*~i})^2
+\lambda_3(\phi^i \phi^{*}_{i})(\hat{\phi}_j \hat{\phi}^{*~j})+
\lambda_4(\phi^i \phi^{*}_{j})
(\hat{\phi}_i \hat{\phi}^{*~j}),
\eeq
This is the most~general renormalizable~form~in four dimensions. In search of a solution of the form of \refeq{powerser} for the REs, among other solutions, one finds in lowest order:
\beq
\begin{split}
g_{Y}&=\hat{g}_{Y}=g~,\\
\lambda_{1}&=\lambda_{2}=\frac{N-1}{N}g^2~,\\
\lambda_{3}&=\frac{1}{2N}g^2~,~
\lambda_{4}=-\frac{1}{2}g^2~,
\end{split}
\eeq
which~corresponds to a $N=1$ SUSY gauge theory. While these remarks do not provide an answer about the relation of reduction of couplings and SUSY, they certainly point to further study in that direction.

\subsection{Reduction of Couplings in N = 1 SUSY Gauge Theories - Partial Reduction}
Consider a chiral,~anomaly~free, $N=1$ globally~supersymmetric
gauge~theory that is based on a group G and has gauge~coupling $g$. The
superpotential of the theory is:
\bea
W&=& \frac{1}{2}\,m_{ij} \,\phi_{i}\,\phi_{j}+
\frac{1}{6}\,C_{ijk} \,\phi_{i}\,\phi_{j}\,\phi_{k}~,
\label{supot0}
\eea
$m_{ij}$ and $C_{ijk}$ are gauge~invariant tensors and
the chiral superfield $\phi_{i}$ belongs to the irreducible~representation  $R_{i}$
of the gauge group. The~renormalization constants~associated with the~superpotential, for preserved SUSY, are:
\begin{align}
\phi_{i}^{0}&=\left(Z^{j}_{i}\right)^{(1/2)}\,\phi_{j}~,~\\
m_{ij}^{0}&=Z^{i'j'}_{ij}\,m_{i'j'}~,~\\
C_{ijk}^{0}&=Z^{i'j'k'}_{ijk}\,C_{i'j'k'}~.
\end{align}
By virtue of the $N=1$ non-renormalization~theorem \cite{Wess:1973kz,Iliopoulos:1974zv,Ferrara:1974fv,Fujikawa:1974ay} there are no~mass and cubic-interaction-term~infinities. Therefore:
\begin{equation}
\begin{split}
Z_{ij}^{i'j'}\left(Z^{i''}_{i'}\right)^{(1/2)}\left(Z^{j''}_{j'}\right)^{(1/2)}
&=\delta_{(i}^{i''}
\,\delta_{j)}^{j''}~,\\
Z_{ijk}^{i'j'k'}\left(Z^{i''}_{i'}\right)^{(1/2)}\left(Z^{j''}_{j'}\right)^{(1/2)}
\left(Z^{k''}_{k'}\right)^{(1/2)}&=\delta_{(i}^{i''}
\,\delta_{j}^{j''}\delta_{k)}^{k''}~.
\end{split}
\end{equation}
The~only surviving~infinities are
the wave~function renormalization~constants $Z^{j}_{i}$, so just one~infinity
per field. The  $\beta$-function of the gauge~coupling $g$ at the one-loop level is given~by
\cite{Parkes:1984dh,West:1984dg,Jones:1985ay,Jones:1984cx,Parkes:1985hh}
\beq
\beta^{(1)}_{g}=\frac{d g}{d t} =
\frac{g^3}{16\pi^2}\left[\,\sum_{i}\,T(R_{i})-3\,C_{2}(G)\right]~,
\label{betag}
\eeq
where $C_{2}(G)$ is the quadratic~Casimir operator of the adjoint~representation of the gauge~group $G$ and $\textrm{Tr}[T^aT^b]=T(R)\delta^{ab}$, where  $T^a$ are the group generators in the appropriate~representation.
The $\beta$-functions of $C_{ijk}$ are related~to the
anomalous~dimension~matrices $\gamma_{ij}$ of the matter~fields as:
\beq
\beta_{ijk} =
 \frac{d C_{ijk}}{d t}~=~C_{ijl}\,\gamma^{l}_{k}+
 C_{ikl}\,\gamma^{l}_{j}+
 C_{jkl}\,\gamma^{l}_{i}~.
\label{betay}
\eeq
The one-loop $\gamma^i_j$ is~given by \cite{Parkes:1984dh}:
\beq
\gamma^{(1)}{}_{j}^{i}=\frac{1}{32\pi^2}\,[\,
C^{ikl}\,C_{jkl}-2\,g^2\,C_{2}(R_{i})\delta^i_j\,],
\label{gamay}
\eeq
where $C^{ijk}=C_{ijk}^{*}$. We take $C_{ijk}$ to be real so that $C_{ijk}^2$  are always~positive. The squares of the couplings are convenient to work with, and the $C_{ijk}$ can be covered~by a single~index $i~(i=1,\cdots,n)$:
\beq
\alpha = \frac{g^2}{4\pi}~,~
\alpha_{i} ~=~ \frac{g_i^2}{4\pi}~.
\label{alfas}
\eeq
Then the evolution of $\alpha$'s in perturbation theory will take~the form
\beq
\begin{split}
\frac{d\alpha}{d t}&=\beta~=~ -\beta^{(1)}\alpha^2+\cdots~,\\
\frac{d\alpha_{i}}{d t}&=\beta_{i}~=~ -\beta^{(1)}_{i}\,\alpha_{i}\,
\alpha+\sum_{j,k}\,\beta^{(1)}_{i,jk}\,\alpha_{j}\,
\alpha_{k}+\cdots~,
\label{eveq}
\end{split}
\eeq
Here, $\cdots$ denotes higher-order contributions~and
$ \beta^{(1)}_{i,jk}=\beta^{(1)}_{i,kj} $.
For the evolution~equations (\ref{eveq}) we investigate the
asymptotic  properties. First, we~define
\cite{Zimmermann:1984sx,Oehme:1985jy,Oehme:1984iz,Cheng:1973nv,Chang:1974bv}
\beq
\tilde{\alpha}_{i} \equiv \frac{\alpha_{i}}{\alpha}~,~i=1,\cdots,n~,
\label{alfat}
\eeq
and~	derive~from \refeq{eveq}
\beq
\begin{split}
\alpha \frac{d \tilde{\alpha}_{i}}{d\alpha} &=
-\tilde{\alpha}_{i}+\frac{\beta_{i}}{\beta}= \left(-1+\frac{\beta^{(1)}_{i}}{\beta^{(1)}}\,\right) \tilde{\alpha}_{i}\\
&
-\sum_{j,k}\,\frac{\beta^{(1)}_{i,jk}}{\beta^{(1)}}
\,\tilde{\alpha}_{j}\, \tilde{\alpha}_{k}+\sum_{r=2}\,
\left(\frac{\alpha}{\pi}\right)^{r-1}\,\tilde{\beta}^{(r)}_{i}(\tilde{\alpha})~,
\label{RE}
\end{split}
\eeq
where $\tilde{\beta}^{(r)}_{i}(\tilde{\alpha})~(r=2,\cdots)$
are power~series of $\tilde{\alpha}$'s and~can be~computed
from the $r^{th}$-loop $\beta$-functions.
We then search for~fixed points $\rho_{i}$ of \refeq{alfat} at $ \alpha
= 0$. We have~to solve the equation
\beq
\left(-1+\frac{\beta ^{(1)}_{i}}{\beta ^{(1)}}\right) \rho_{i}
-\sum_{j,k}\frac{\beta ^{(1)}_{i,jk}}{\beta ^{(1)}}
\,\rho_{j}\, \rho_{k}=0~,
\label{fixpt}
\eeq
assuming fixed~points of the~form
\beq
\rho_{i}=0~\mbox{for}~ i=1,\cdots,n'~;~
\rho_{i} ~>0 ~\mbox{for}~i=n'+1,\cdots,n~.
\eeq
Next, we treat $\tilde{\alpha}_{i}$ with $i \leq n'$
as small~perturbations  to the
undisturbed~system (defined by~setting
$\tilde{\alpha}_{i}$  with $i \leq n'$ equal~to zero).
It is possible~to verify the
existence~of the unique power~series solution of the reduction~equations (\ref{RE}) to all orders already at one-loop level \cite{Zimmermann:1984sx,Oehme:1984yy,Oehme:1985jy,Oehme:1984iz}:
\beq
\tilde{\alpha}_{i}=\rho_{i}+\sum_{r=2}\rho^{(r)}_{i}\,
\alpha^{r-1}~,~i=n'+1,\cdots,n~.
\label{usol}
\eeq
 These~are RGI~relations among parameters, and preserve formally
perturbative~renormalizability.
So, in the~undisturbed system~there is only one~independent
parameter, the primary coupling $\alpha$.

The nonvanishing $\tilde{\alpha}_{i}$
with $i \leq n'$ cause small perturbations that enter in a way that the reduced~couplings
($\tilde{\alpha}_{i}$  with $i > n'$) become functions both of
$\alpha$ and $\tilde{\alpha}_{i}$  with $i \leq n'$.
Investigating such systems with partial reduction is very convenient to work with the following PDEs:
\beq
\begin{split}
\left\{ \tilde{\beta}\,\frac{\partial}{\partial\alpha}
+\sum_{a=1}^{n'}\,
\tilde{\beta_{a}}\,\frac{\partial}{\partial\tilde{\alpha}_{a}}\right\}~
\tilde{\alpha}_{i}(\alpha,\tilde{\alpha})
&=\tilde{\beta}_{i}(\alpha,\tilde{\alpha})~,\\
\tilde{\beta}_{i(a)}~=~\frac{\beta_{i(a)}}{\alpha^2}
-\frac{\beta}{\alpha^{2}}~\tilde{\alpha}_{i(a)}
&,\qquad
\tilde{\beta}~\equiv~\frac{\beta}{\alpha}~.
\end{split}
\eeq
These equations are~equivalent~to the REs (\ref{RE}), where, in order to~avoid any confusion, we let
$a,b$ run from $1$ to $n'$ and $i,j$ from $n'+1$ to $n$. Then, we search for solutions~of the~form
\beq
\tilde{\alpha}_{i}=\rho_{i}+
\sum_{r=2}\,\left(\frac{\alpha}{\pi}\right)^{r-1}\,f^{(r)}_{i}
(\tilde{\alpha}_{a})~,~i=n'+1,\cdots,n~,
\label{algeq}
\eeq
where $ f^{(r)}_{i}(\tilde{\alpha}_{a})$ are power~series of $\tilde{\alpha}_{a}$. 
The requirement that~in the~limit of~vanishing
perturbations~we obtain~the undisturbed~solutions (\ref{usol})
\cite{Kubo:1988zu,Zimmermann:1993ei} suggests this type of solutions. Once more, one can obtain  the conditions for uniqueness of $ f^{(r)}_{i}$ in~terms of~the lowest~order
coefficients.

\subsection{Reduction of Dimension-1 and -2 Parameters}\label{sec:dimful}

The extension of reduction of couplings to massive~parameters is not
straightforward, since the technique was originally aimed at massless~theories on~the basis~of
the~Callan-Symanzik equation \cite{Zimmermann:1984sx,Oehme:1984yy}. Many requirements have to be met, such  as the normalization conditions imposed on irreducible Green's functions
\cite{Piguet:1989pc}, etc. Significant progress has been made towards this goal, starting from~\cite{Kubo:1996js}, where, as an assumption, a mass-independent renormalization scheme 
renders all RG functions only trivially dependent on dimensional~parameters. Mass parameters can
then be introduced similarly to couplings. 

This  was
justified~later  \cite{Breitenlohner:2001pp,Zimmermann:2001pq}, where it was demonstrated that, apart from dimensionless parameters,
pole~masses and gauge couplings, the model can also include couplings carrying a dimension and masses.
To simplify the~analysis, we follow \citere{Kubo:1996js} and use a~mass-independent renormalization~scheme as well.

Consider a renormalizable~theory that contains $(N + 1)$
dimension-0 couplings, $\left(\hat g_0,\hat g_1, ...,\hat g_N\right)$,
 $L$ parameters with~mass dimension-1, $\left(\hat h_1,...,\hat h_L\right)$,
and  $M$ parameters~with mass dimension-2, $\left(\hat m_1^2,...,\hat m_M^2\right)$.
The renormalized~irreducible vertex~function $\Gamma$ satisfies the RG
equation
\beq
\label{RGE_OR_1}
\mathcal{D}\Gamma\left[\Phi's;\hat g_0,\hat g_1, ...,\hat g_N;\hat h_1,...,\hat h_L;\hat m_1^2,...,\hat m_M^2;\mu\right]=0~,
\eeq
with
\beq
\label{RGE_OR_2}
\mathcal{D}=\mu\frac{\partial}{\partial \mu}+
\sum_{i=0}^N \beta_i\frac{\partial}{\partial \hat g_i}+
\sum_{a=1}^L \gamma_a^h\frac{\partial}{\partial \hat h_a}+
\sum_{\alpha=1}^M \gamma_\alpha^{m^2}\frac{\partial}{\partial \hat m_\alpha ^2}+
\sum_J \Phi_I\gamma^{\phi I}_{\,\,\,\, J}\,\frac{\delta}{\delta\Phi_J}~,
\eeq
where  $\beta_i$ are the $\beta$-functions of the dimensionless~couplings $g_i$ and $\Phi_I$  are
the 
 matter~fields. The mass, trilinear~coupling and wave~function anomalous~dimensions,
respectively, are denoted by 
$\ga_\alpha^{m^2}$, $\ga_a^h$ and $\ga^{\phi I}_{\,\,\,\, J}$
 and $\mu$ denotes the energy scale.
For a mass-independent renormalization~scheme, the $\gamma$'s are given by
\beq
\label{gammas}
\begin{split}
\gamma^h_a&=\sum_{b=1}^L\gamma_a^{h,b}(g_0,g_1,...,g_N)\hat h_b,\\
\gamma_\alpha^{m^2}&=\sum_{\beta=1}^M \gamma_\alpha^{m^2,\beta}(g_0,g_1,...,g_N)\hat m_\beta^2+
\sum_{a,b=1}^L \gamma_\alpha^{m^2,ab}(g_0,g_1,...,g_N)\hat h_a\hat h_b~.
\end{split}
\eeq
The $\gamma_a^{h,b}$, $\gamma_\alpha^{m^2,\beta}$ and $\gamma_\alpha^{m^2,ab}$ are power~series of the (dimensionless)
$g$'s.

\vspace{0.35cm}

\noindent We search for~a reduced~theory where
\[
g\equiv g_0,\qquad h_a\equiv \hat h_a\quad \textrm{for $1\leq a\leq P$},\qquad
m^2_\alpha\equiv\hat m^2_\alpha\quad \textrm{for $1\leq \alpha\leq Q$}
\]
are independent parameters. The reduction of the rest of the parameters, namely
\beq
\label{reduction}
\begin{split}
\hat g_i &= \hat g_i(g), \qquad (i=1,...,N),\\
\hat h_a &= \sum_{b=1}^P f_a^b(g)h_b, \qquad (a=P+1,...,L),\\
\hat m^2_\alpha &= \sum_{\beta=1}^Q e^\beta_\alpha(g)m^2_\beta + \sum_{a,b=1}^P k^{ab}_\alpha(g)h_ah_b,
\qquad (\alpha=Q+1,...,M)
\end{split}
\eeq
is~consistent with~the RGEs (\ref{RGE_OR_1},\ref{RGE_OR_2}). The~following relations~should be~satisfied
\beq
\label{relation}
\begin{split}
\beta_g\,\frac{\partial\hat g_i}{\partial g} &= \beta_i,\qquad (i=1,...,N),\\
\beta_g\,\frac{\partial \hat h_a}{\partial g}+\sum_{b=1}^P \gamma^h_b\,\frac{\partial\hat h_a}{\partial h_b} &= \gamma^h_a,\qquad (a=P+1,...,L),\\
\beta_g\,\frac{\partial\hat m^2_\alpha}{\partial g} +\sum_{a=1}^P \gamma_a^h\,\frac{\partial\hat m^2_\alpha}{\partial h_a} +  \sum_{\beta=1}^Q \gamma_\beta ^{m^2}\,\frac{\partial\hat m_\alpha^2}{\partial m_\beta^2} &= \gamma_\alpha^{m^2}, \qquad (\alpha=Q+1,...,M).
\end{split}
\eeq
Using Eqs. (\ref{gammas}) and~(\ref{reduction}), they reduce~to
\beq
\label{relation_2}
\begin{split}
&\beta_g\,\frac{df^b_a}{dg}+ \sum_{c=1}^P f^c_a\left[\gamma^{h,b}_c + \sum_{d=P+1}^L \gamma^{h,d}_c f^b_d\right] -\gamma^{h,b}_a - \sum_{d=P+1}^L \gamma^{h,d}_a f^b_d=0,\\
&\hspace{8.6cm} (a=P+1,...,L;\, b=1,...,P),\\
&\beta_g\,\frac{de^\beta_\alpha}{dg} + \sum_{\gamma=1}^Q e^\gamma_\alpha\left[\gamma_\gamma^{m^2,\beta} +
\sum _{\delta=Q+1}^M\gamma_\gamma^{m^2,\delta} e^\beta_\delta\right]-\gamma_\alpha^{m^2,\beta} -
\sum_{\delta=Q+1}^M \gamma_\alpha^{m^2,d}e^\beta_\delta =0,\\
&\hspace{8.3cm} (\alpha=Q+1,...,M ;\, \beta=1,...,Q),\\
&\beta_g\,\frac{dk_\alpha^{ab}}{dg}
+ 2\sum_{c=1}^P \left(\gamma_c^{h,a} + \sum_{d=P+1}^L \gamma_c^{h,d} f_d^a\right)k_\alpha^{cb}
+\sum_{\beta=1}^Q e^\beta_\alpha\left[\gamma_\beta^{m^2,ab} + \sum_{c,d=P+1}^L \gamma_\beta^{m^2,cd}f^a_cf^b_d \right.\\
&\left. +2\sum_{c=P+1}^L \gamma_\beta^{m^2,cb}f^a_c + \sum_{\delta=Q+1}^M \gamma_\beta^{m^2,d} k_\delta^{ab}\right]- \left[\gamma_\alpha^{m^2,ab}+\sum_{c,d=P+1}^L \gamma_\alpha^{m^2,cd}f^a_c f^b_d\right.\\
&\left. +2 \sum_{c=P+1}^L \gamma_\alpha^{m^2,cb}f^a_c + \sum_{\delta=Q+1}^M \gamma_\alpha^{m^2,\delta}k_\delta^{ab}\right]=0,\\
&\hspace{8cm} (\alpha=Q+1,...,M;\, a,b=1,...,P)~.
\end{split}
\eeq
The~above relations ensure~that the~irreducible vertex~function of~the reduced~theory
\beq
\label{Green}
\begin{split}
\Gamma_R&\left[\Phi\textrm{'s};g;h_1,...,h_P; m_1^2,...,m_Q^2;\mu\right]\equiv\\
&\Gamma \left[  \Phi\textrm{'s};g,\hat g_1(g)...,\hat g_N(g);
h_1,...,h_P,\hat h_{P+1}(g,h),...,\hat h_L(g,h);\right.\\
& \left.  \qquad\qquad\qquad m_1^2,...,m^2_Q,\hat m^2_{Q+1}(g,h,m^2),...,\hat m^2_M(g,h,m^2);\mu\right]
\end{split}
\eeq
has~the same~renormalization group~flow as~the original~one.

Assuming a perturbatively~renormalizable reduced~theory, the functions
$\hat g_i$, $f^b_a$, $e^\beta_\alpha$ and $k_\alpha^{ab}$ are
expressed~as power~series in~the primary~coupling:
\beq
\label{pert}
\begin{split}
\hat g_i & = g\sum_{n=0}^\infty \rho_i^{(n)} g^n,\qquad
f_a^b  =  g \sum_{n=0}^\infty \eta_a^{b(n)} g^n,\\
e^\beta_\alpha & = \sum_{n=0}^\infty \xi^{\beta(n)}_\alpha g^n,\qquad
k_\alpha^{ab}=\sum_{n=0}^\infty \chi_\alpha^{ab(n)} g^n.
\end{split}
\eeq
These expansion~coefficients are found by~inserting the above power~series into Eqs. (\ref{relation}), (\ref{relation_2}) and~requiring the~equations to~be satisfied~at each~order of~$g$. It is not trivial to have a unique power series solution; it depends both on the theory and  the choice of independent couplings.

If there are no independent dimension-1 parameters ($\hat h$), their reduction becomes
\[
\hat h_a = \sum_{b=1}^L f_a^b(g)M,
\]
where $M$ is a dimension-1 parameter (i.e. a gaugino~mass, corresponding to the independent gauge coupling). If there
are no independent dimension-2
parameters ($\hat m^2$), their reduction~takes the form
\[
\hat m^2_a=\sum_{b=1}^M e_a^b(g) M^2.
\]

\subsection{Reduction of Couplings of Soft Breaking Terms in $N=1$ SUSY Theories}\label{sec:RCN=1}

The reduction of dimensionless couplings was~extended \cite{Kubo:1996js,Jack:1995gm} to~the SSB dimensionful~parameters of $N = 1$
supersymmetric~theories. It was also found \cite{Kawamura:1997cw,Kobayashi:1997qx} that
soft scalar masses satisfy~a universal~sum rule.\\
We consider the superpotential (\ref{supot0})
\be
W= \frac{1}{2}\,\mu^{ij} \,\Phi_{i}\,\Phi_{j}+
\frac{1}{6}\,C^{ijk} \,\Phi_{i}\,\Phi_{j}\,\Phi_{k}~,
\label{supot}
\ee
and the SSB Lagrangian 
\be
-{\cal L}_{\rm SSB} =
\frac{1}{6} \,h^{ijk}\,\phi_i \phi_j \phi_k
+
\frac{1}{2} \,b^{ij}\,\phi_i \phi_j
+
\frac{1}{2} \,(m^2)^{j}_{i}\,\phi^{*\,i} \phi_j+
\frac{1}{2} \,M\,\lambda_i \lambda_i+\mbox{h.c.}
\label{supot_l}
\ee
The $\phi_i$'s are~the scalar~parts of chiral~superfields $\Phi_i$, $\lambda$ are gauginos
and $M$ the unified gaugino mass.

The one-loop gauge $\beta$-function (\ref{betag}) is given by
\cite{Parkes:1984dh,West:1984dg,Jones:1985ay,Jones:1984cx,Parkes:1985hh}
\bea
\beta^{(1)}_{g}=\frac{d g}{d t} =
  \frac{g^3}{16\pi^2}\,\left[\,\sum_{i}\,T(R_{i})-3\,C_{2}(G)\,\right]~,
\eea
whereas the one-loop $C_{ijk}$'s $\beta$-function (\ref{betay}) is~given by
\be
\beta_C^{ijk} =
  \frac{d C_{ijk}}{d t}~=~C_{ijl}\,\gamma^{l}_{k}+
  C_{ikl}\,\gamma^{l}_{j}+
  C_{jkl}\,\gamma^{l}_{i}~,
\ee
and the (one-loop) anomalous dimension $\gamma^{(1)}\,^i_j$ of a chiral~superfield (\ref{gamay}) is
\be
\gamma^{(1)}\,^i_j=\frac{1}{32\pi^2}\,\left[\,
C^{ikl}\,C_{jkl}-2\,g^2\,C_{2}(R_{i})\delta^i_j\,\right]~.
\label{allolabel}
\ee 
Then the $N = 1$ non-renormalization~theorem \cite{Wess:1973kz,Iliopoulos:1974zv,Fujikawa:1974ay} guarantees that the $\beta$-functions of $C_{ijk}$ are expressed in terms of the anomalous dimensions.\\
We make the assumption that the REs admit~power series~solutions:
\be
C^{ijk} = g\,\sum_{n=0}\,\rho^{ijk}_{(n)} g^{2n}~.
\label{Yg}
\ee
Since we want to obtain higher-loop results instead~of knowledge~of explicit~$\beta$-functions, we require relations among $\beta$-functions. The spurion technique
\cite{Fujikawa:1974ay,Delbourgo:1974jg,Salam:1974pp,Grisaru:1979wc,Girardello:1981wz} gives all-loop relations among SSB $\beta$-functions
\cite{Yamada:1994id,Kazakov:1997nf,Jack:1997pa,Hisano:1997ua,Jack:1997eh,Avdeev:1997vx,Kazakov:1998uj,Karch:1998qa}:
\begin{align}
\beta_M &= 2{\cal O}\left(\frac{\beta_g}{g}\right)~,
\label{betaM}\\
\beta_h^{ijk}&=\gamma^i_l h^{ljk}+\gamma^j_l h^{ilk}
+\gamma^k_l h^{ijl}\non\\
&\,-2\left(\gamma_1\right)^i_l C^{ljk}
-2\left(\gamma_1\right)^j_l C^{ilk}-2\left(\gamma_1\right)^k_l C^{ijl}~,\label{betah}\\
(\beta_{m^2})^i_j &=\left[ \Delta
+ X \frac{\partial}{\partial g}\right]\gamma^i_j~,
\label{betam2}
\end{align}
where
\begin{align}
{\cal O} &=\left(Mg^2\frac{\partial}{\partial g^2}
-h^{lmn}\frac{\partial}{\partial C^{lmn}}\right)~,
\label{diffo}\\
\Delta &= 2{\cal O}{\cal O}^* +2|M|^2 g^2\frac{\partial}
{\partial g^2} +\tilde{C}_{lmn}
\frac{\partial}{\partial C_{lmn}} +
\tilde{C}^{lmn}\frac{\partial}{\partial C^{lmn}}~,\\
(\gamma_1)^i_j&={\cal O}\gamma^i_j,\\
\tilde{C}^{ijk}&=
(m^2)^i_l C^{ljk}+(m^2)^j_l C^{ilk}+(m^2)^k_l C^{ijl}~.
\label{tildeC}
\end{align}

Assuming (following \cite{Jack:1997pa}) that~the relation~among couplings
\be
h^{ijk} = -M (C^{ijk})'
\equiv -M \frac{d C^{ijk}(g)}{d \ln g}~,
\label{h2}
\ee
is RGI and  the~use of~the all-loop gauge $\beta$-function of 
\cite{Novikov:1983ee,Novikov:1985rd,Shifman:1996iy}
\be
\beta_g^{\rm NSVZ} =
\frac{g^3}{16\pi^2}
\left[ \frac{\sum_l T(R_l)(1-\gamma_l /2)
-3 C_2(G)}{ 1-g^2C_2(G)/8\pi^2}\right]~,
\label{bnsvz}
\ee
we are led to an all-loop RGI sum rule \cite{Kobayashi:1998jq} (assuming $(m^2)^i_j=m^2_j\delta^i_j$),
\begin{equation}
\begin{split}
m^2_i+m^2_j+m^2_k &=
|M|^2 \left\{~
\frac{1}{1-g^2 C_2(G)/(8\pi^2)}\frac{d \ln C^{ijk}}{d \ln g}
+\frac{1}{2}\frac{d^2 \ln C^{ijk}}{d (\ln g)^2}~\right\}\\
& \qquad\qquad +\sum_l
\frac{m^2_l T(R_l)}{C_2(G)-8\pi^2/g^2}
\frac{d \ln C^{ijk}}{d \ln g}~.
\label{sum2}
\end{split}
\end{equation}
It is worth noting that the all-loop result of \refeq{sum2} coincides~with
the~superstring result~for the~finite case~in a certain~class of orbifold~models
\cite{Ibanez:1992hc,Brignole:1995fb,Kobayashi:1997qx}
if $\frac{d \ln C^{ijk}}{d \ln g}=1$~\cite{Mondragon:1993tw}.

As mentioned above, the~all-loop results~on the~SSB $\beta$-functions, Eqs.(\ref{betaM})-(\ref{tildeC}),
lead~to all-loop~RGI relations. We assume:\\
(a) the~existence of~an RGI~surface on~which $C = C(g)$, or~equivalently that~the expression
\be
\label{Cbeta}
\frac{dC^{ijk}}{dg} = \frac{\beta^{ijk}_C}{\beta_g}
\ee
holds  (i.e. reduction of couplings is possible)\\
(b) the~existence of~a RGI~surface on~which
\be
\label{h2NEW}
h^{ijk} = - M \frac{dC(g)^{ijk}}{d\ln g}
\ee
holds to~all orders.\\
Then it can be proven \cite{Jack:1999aj,Kobayashi:1998iaa} that the relations that follow are all-loop RGI (note that in
both assumptions we do not rely on specific solutions of these equations)
\begin{align}
M &= M_0~\frac{\beta_g}{g} ,  \label{M-M0} \\
h^{ijk}&=-M_0~\beta_C^{ijk},  \label{hbeta}  \\
b^{ij}&=-M_0~\beta_{\mu}^{ij},\label{bij}\\
(m^2)^i_j&= \frac{1}{2}~|M_0|^2~\mu\frac{d\gamma^i{}_j}{d\mu},
\label{scalmass}
\end{align}
where $M_0$~is an~arbitrary reference~mass scale~to be~specified shortly. Assuming
\be
C_a\frac{\partial}{\partial C_a}
= C_a^*\frac{\partial}{\partial C_a^*} \label{dc/dc}
\ee
for an RGI~surface $F(g,C^{ijk},C^{*ijk})$ we are led to
\begin{equation}
\label{F}
\frac{d}{dg} = \left(\frac{\partial}{\partial g} + 2\frac{\partial}{\partial C}\,\frac{dC}{dg}\right)
= \left(\frac{\partial}{\partial g} + 2 \frac{\beta_C}{\beta_g}
\frac{\partial}{\partial C} \right)\, ,
\end{equation}
where \refeq{Cbeta} was used. Let us now consider~the partial~differential operator ${\cal O}$ in
\refeq{diffo} which (assuming \refeq{h2}), becomes
\be
{\cal O} = \frac{1}{2}M\frac{d}{d\ln g}\, 
\ee
and $\beta_M$, given~in \refeq{betaM}, becomes
\be
\beta_M = M\frac{d}{d\ln g} \big( \frac{\beta_g}{g}\big) ~, \label{betaM2}
\ee
which~by integration~provides us \cite{Karch:1998qa,Jack:1999aj} with~the
generalized, i.e. including Yukawa couplings, all-loop RGI Hisano - Shifman relation \cite{Hisano:1997ua}
\be
 M = \frac{\beta_g}{g} M_0~. \nn
\ee
$M_0$ is the~integration constant~and can~be associated~to the unified gaugino mass $M$ (of an assumed covering GUT), or to~the gravitino~mass $m_{3/2}$ in
a supergravity~framework. Therefore, \refeq{M-M0} becomes the
all-loop RGI \refeq{M-M0}.  $\beta_M$, using
Eqs.(\ref{betaM2}) and (\ref{M-M0}) can be written as follows:
\be \beta_M =
M_0\frac{d}{dt} (\beta _g/g)~.
\ee
Similarly
\be (\gamma_1)^i_j =
{\cal O} \gamma^i_j = \frac{1}{2}~M_0~\frac{d
  \gamma^i_j}{dt}~. \label{gammaO}
\ee
Next, from Eq.(\ref{h2}) and Eq.(\ref{M-M0}) we get
\be
 h^{ijk} = - M_0 ~\beta_C^{ijk}~,  \label{hm32}
\ee
while $\beta^{ijk}_h$, using Eq.(\ref{gammaO}), becomes \cite{Jack:1999aj}
\be
  \beta_h^{ijk} = - M_0~\frac{d}{dt} \beta_C^{ijk},
\ee
which shows that \refeq{hm32} is  RGI to all loops. \refeq{bij} can similarly be shown to~be all-loop~RGI as well.

Finally, it is important to note that,~under the assumptions (a)
and (b), the sum rule of \refeq{sum2} has been proven
\cite{Kobayashi:1998jq} to be RGI to all loops, which (using \refeq{M-M0}) generalizes \refeq{scalmass} for application in cases with non-universal soft~scalar masses, a
necessary ingredient in the models that will be examined in the next Sections. Another important point to note is the use of \refeq{M-M0}, which, in the case of product gauge groups (as in the MSSM), takes the form 
\be
M_i=\frac{\beta_{g_i}}{g_i}M_0~,
\ee
where $i=1,2,3$ denotes each gauge group, and will be used in the Reduced MSSM case.


\section{Finiteness in N=1 SUSY Gauge Theories}\label{se:fin}

 We start by considering a chiral,~anomaly free, $N=1$ globally~supersymmetric
gauge~theory with gauge~group G and $g$ the theory's coupling constant.
Again, the theory's superpotential is~given by \refeq{supot0}.
Because of the $N=1$ non-renormalization~theorem, the one-loop $\beta$-function is given~by \refeq{betag},
the $\beta$-function of $C_{ijk}$ by \refeq{betay}
and the one-loop anomalous dimensions of the chiral superfields by \refeq{gamay}.

\noindent It is obvious from Eqs.~(\ref{betag}) and  (\ref{gamay}) that
 all  one-loop $\beta$-functions of the  theory vanish if
$\beta_g^{(1)}$ and  $\gamma^{(1)}{}_{j}^{i}$ vanish:
\begin{align}
\sum _i T(R_{i})& = 3 C_2(G) \,,
\label{1st}     \\
 C^{ikl} C_{jkl} &= 2\delta ^i_j g^2  C_2(R_i)~.
\label{2nd}
\end{align}

\noindent In \cite{Rajpoot:1984zq} one can find
the finiteness conditions for $N=1$ theories with $SU(N)$ gauge
symmetry, while \cite{Rajpoot:1985aq} discusses the requirements of anomaly-free and no-charge renormalization. 
Remarkably, the  conditions (\ref{1st},\ref{2nd})
are necessary~and  sufficient~for finiteness at the  two-loop level as well
\cite{Parkes:1984dh,West:1984dg,Jones:1985ay,Jones:1984cx,Parkes:1985hh}.

In the case of soft SUSY breaking, requiring
finiteness in the  one-loop SSB sector imposes additional
constraints among soft terms \cite{Jones:1984cu}.  Again, the one-loop SSB finiteness conditions  are enough to render the~soft sector~two-loop~finite \cite{Jack:1994kd}.

The above finiteness conditions impose considerable restrictions on the choice of irreducible representations
(irreps)
$R_i$ for a~given
group $G$ as well as the  Yukawa couplings.  These conditions cannot be
applied to the MSSM, because the $U(1)$ gauge group is not compatible with condition
(\ref{1st}), since $C_2[U(1)]=0$.  This points to the grand unified
level, with the MSSM just being the low-energy~theory.

Additionally,  one(two)-loop finiteness causes 
SUSY to break only softly. Since gauge singlets are not acceptable, due to the condition given in \refeq{2nd} ($C_2(1)=0$, i.e. singlets do not couple to the theory),
F-type~spontaneous symmetry~breaking \cite{O'Raifeartaigh:1975pr}
terms~are incompatible~with finiteness, as~well as D-type
\cite{Fayet:1974jb} spontaneous~breaking which~requires the~existence
of a $U(1)$ gauge~group.

One can see that conditions (\ref{1st},\ref{2nd}) impose relations between the gauge and  Yukawa
sector.  Imposing such relations, that make the  parameters
mutually~dependent at a given renormalization~point, is trivial.  What
is not~trivial is to~guarantee that~relations leading~to a reduction
of the~couplings hold~at any~renormalization~point.  As explained
(see \refeq{Cbeta}),
the~necessary and~sufficient condition is~to
require~that such relations~are solutions~to the  REs
\beq \beta _g
\frac{d C_{ijk}}{dg} = \beta _{ijk}
\label{redeq2}
\eeq
and hold~at all~orders. It is reminded that the existence of
all-order power~series solutions~to (\ref{redeq2}) can~be decided~at
one-loop level.

Concerning higher loop orders,
a theorem
\cite{Lucchesi:1987he,Lucchesi:1987ef} exists that~states the  necessary~and
sufficient~conditions to achieve~all-loop finiteness for an $N=1$~SUSY theory.  It relies on the  structure of the~supercurrent in an
$N=1$~SUSY theory
\cite{Ferrara:1974pz,Piguet:1981mu,Piguet:1981mw}, and  on the
non-renormalization~properties of $N=1$ chiral~anomalies
\cite{Lucchesi:1987he,Lucchesi:1987ef,Piguet:1986td,Piguet:1986pk,Ensign:1987wy}.
Details and further~discussion can be found in
\cite{Lucchesi:1987he,Lucchesi:1987ef, Piguet:1986td,Piguet:1986pk,Ensign:1987wy,Lucchesi:1996ir,Piguet:1996mx}
Following \cite{Piguet:1996mx} we briefly discuss the  proof.

Consider~an $N=1$ SUSY~gauge theory, with~simple Lie~group
$G$.  The~content of this theory~is given at the~classical level~by
the matter supermultiplets~$S_i$, which~contain a scalar~field
$\phi_i$ and  a Weyl~spinor $\psi_{ia}$, and  the~vector supermultiplet
$V_a$, which~contains a gauge~vector~field $A_{\mu}^a$ and  a gaugino
Weyl~spinor $\lambda^a_{\alpha}$.

Let us~first recall~certain facts~about the~theory:

\noindent (1)  A~massless $N=1$ SUSY~theory is invariant~under a $U(1)$ chiral~transformation $R$ under~which the  various~fields
transform~as follows
\beq
\begin{split}
A'_{\mu}&=A_{\mu},~~\lambda '_{\alpha}=\exp({-i\theta})\lambda_{\alpha}\\
\phi '&= \exp({-i\frac{2}{3}\theta})\phi,~~\psi_{\alpha}'= \exp({-i\frac{1}
    {3}\theta})\psi_{\alpha},~\cdots
\end{split}
\eeq
The corresponding~axial Noether~current $J^{\mu}_R(x)$ is
\beq
J^{\mu}_R(x)=\bar{\lambda}\gamma^{\mu}\gamma^5\lambda + \cdots
\label{noethcurr}
\eeq
is conserved~classically, while~in the  quantum~case is violated~by the
axial~anomaly
\beq
\partial_{\mu} J^{\mu}_R =
r\left(\epsilon^{\mu\nu\sigma\rho}F_{\mu\nu}F_{\sigma\rho}+\cdots\right).
\label{anomaly}
\eeq

From its known~topological origin~in ordinary~gauge theories
\cite{AlvarezGaume:1983cs,Bardeen:1984pm,Zumino:1983rz}, one would~expect the  axial~vector~current
$J^{\mu}_R$ to~satisfy the  Adler-Bardeen~theorem  and
receive~corrections only~at the  one-loop~level.  Indeed~it has~been
shown~that the~same non-renormalization~theorem holds~also in
SUSY~theories \cite{Piguet:1986td,Piguet:1986pk,Ensign:1987wy}.  Therefore
\beq
r=\hbar \beta_g^{(1)}.
\label{r}
\eeq

\noindent (2)  The  massless~theory we~consider is scale~invariant at
the~classical~level and, in~general, there is a scale~anomaly due to
radiative~corrections.  The  scale~anomaly appears~in the  trace~of the
energy~momentum~tensor $T_{\mu\nu}$, which is traceless~classically.
It has the~form
\beq
T^{\mu}_{\mu} = \beta_g F^{\mu\nu}F_{\mu\nu} +\cdots
\label{Tmm}
\eeq

\noindent (3)  Massless, $N=1$ SUSY~gauge theories~are
classically~invariant under the  supersymmetric~extension of the
conformal~group -- the  superconformal~group.  Examining the
superconformal~algebra, it~can be~seen that~the  subset~of
superconformal~transformations~consisting of translations,
SUSY~transformations, and~axial $R$~transformations is~closed
under~SUSY, i.e. these~transformations form~a representation
of~SUSY.  It~follows that~the  conserved~currents
corresponding~to these~transformations make~up a~supermultiplet
represented~by an~axial vector~superfield called~the  supercurrent~$J$,
\beq
J \equiv \left\{ J'^{\mu}_R, ~Q^{\mu}_{\alpha}, ~T^{\mu}_{\nu} , ... \right\},
\label{J}
\eeq
where $J'^{\mu}_R$ is~the  current~associated to~R-invariance,
$Q^{\mu}_{\alpha}$ is~the  one~associated to SUSY~invariance,
and $T^{\mu}_{\nu}$ the~one associated~to translational~invariance
(energy-momentum~tensor).

The~anomalies of the~R-current $J'^{\mu}_R$, the~trace
anomalies~of the
SUSY~current, and~the  energy-momentum~tensor, form~also
a second~supermultiplet, called the~supertrace~anomaly
\[
S =\left\{ Re~ S, ~Im~ S,~S_{\alpha}\right\}=
\left\{T^{\mu}_{\mu},~\partial _{\mu} J'^{\mu}_R,~\sigma^{\mu}_{\alpha
  \dot{\beta}} \bar{Q}^{\dot\beta}_{\mu}~+~\cdots \right\}
\]
where~$T^{\mu}_{\mu}$ is~given in \refeq{Tmm} and
\begin{align}
\partial _{\mu} J'^{\mu}_R &~=~\beta_g\epsilon^{\mu\nu\sigma\rho}
F_{\mu\nu}F_{\sigma \rho}+\cdots\\
\sigma^{\mu}_{\alpha \dot{\beta}} \bar{Q}^{\dot\beta}_{\mu}&~=~\beta_g
\lambda^{\beta}\sigma ^{\mu\nu}_{\alpha\beta}F_{\mu\nu}+\cdots
\end{align}

\noindent (4) It~is important to~note that~the Noether~current defined~in (\ref{noethcurr}) is~not the~same as~the
current~associated to~R-invariance~that appears~in the~supercurrent
$J$ in (\ref{J}), but~they coincide~in the tree~approximation.
So~starting from~a unique~classical Noether~current
$J^{\mu}_{R(class)}$,  the~Noether
current $J^{\mu}_R$ is~defined as the~quantum extension~of
$J^{\mu}_{R(class)}$ which~allows for~the
validity~of the~non-renormalization~theorem.  On~the  other~hand,
$J'^{\mu}_R$, is~defined to~belong to~the  supercurrent $J$,
together~with the~energy-momentum~tensor.  The  two~requirements
cannot~be fulfilled~by a single~current operator~at the  same~time.

Although~the  Noether~current $J^{\mu}_R$ which~obeys (\ref{anomaly})
and~the  current $J'^{\mu}_R$ belonging~to the  supercurrent~multiplet
$J$ are~not the~same, there~is a~relation
\cite{Lucchesi:1987he,Lucchesi:1987ef} between quantities~associated
with~them
\beq
r=\beta_g(1+x_g)+\beta_{ijk}x^{ijk}-\gamma_Ar^A~,
\label{rbeta}
\eeq
where $r$ is~given in \refeq{r}.  The  $r^A$ are~the
non-renormalized~coefficients of~the anomalies~of the ~Noether currents~associated to the  chiral~invariances of the~superpotential, and  --like $r$-- are~strictly
one-loop~quantities. The  $\gamma_A$'s are~linear
combinations~of the  anomalous~dimensions of~the  matter~fields, and
$x_g$, and  $x^{ijk}$ are~radiative correction~quantities.
The~structure of \refeq{rbeta} is~independent of~the
renormalization~scheme.

One-loop~finiteness, i.e. vanishing~of the  $\beta$-functions~at one loop,
implies~that the  Yukawa~couplings $\lambda_{ijk}$ must~be functions of~the gauge~coupling $g$. To~find a similar~condition to all~orders it
is~necessary and~sufficient for~the  Yukawa~couplings to~be a~formal
power~series in $g$, which~is solution~of the~REs (\ref{redeq2}).

\bigskip
We~can now~state the~theorem for~all-order vanishing
$\beta$-functions \cite{Lucchesi:1987he}.

\noindent
{\bf Theorem:}\\
Consider~an $N=1$ SUSY~Yang-Mills theory, with simple~gauge
group. If~the  following~conditions are~satisfied
\begin{enumerate}
\item There~is no gauge~anomaly.
\item The~gauge $\beta$-function vanishes~at one loop
  \beq
  \beta^{(1)}_g = 0 =\sum_i T(R_{i})-3\,C_{2}(G).
  \eeq
\item There~exist solutions~of the~form
  \beq
  C_{ijk}=\rho_{ijk}g,~\qquad \rho_{ijk}\in\complex
  \label{soltheo}
  \eeq
to~the   conditions~of vanishing~one-loop matter~fields anomalous~dimensions
\beq
  \gamma^{(1)}{}_{j}^{i}~=~0
  =\frac{1}{32\pi^2}~[ ~
  C^{ikl}\,C_{jkl}-2~g^2~C_{2}(R)\delta_j^i ].
\eeq
\item These~solutions are~isolated and~non-degenerate when~considered
  as~solutions of~vanishing one-loop~Yukawa $\beta$-functions:
   \beq
   \beta_{ijk}=0.
   \eeq
\end{enumerate}
Then, each~of the~solutions (\ref{soltheo}) can~be uniquely~extended
to~a formal~power series~in $g$, and~the  associated~super Yang-Mills~models depend~on the~single coupling~constant $g$ with~a $\beta$-function
which~vanishes at all orders.

\bigskip

Important note:
The~requirement of~isolated and~non-degenerate
solutions~guarantees the~existence of~a unique~formal power~series solution~to the~reduction
equations.
The~vanishing of~the  gauge~$\beta$-function at~one loop,
$\beta_g^{(1)}$, is~equivalent to~the
vanishing~of the~R-current~anomaly (\ref{anomaly}).  The~vanishing of~the anomalous~dimensions at~one loop~implies the~vanishing of~the  Yukawa~couplings
$\beta$-functions at~that order.  It~also implies~the  vanishing~of the~chiral anomaly~coefficients $r^A$.  This~last property~is a~necessary
condition~for having $\beta$-functions vanishing~at all~orders.\footnote{There~is an~alternative way~to find~finite theories
\cite{Ermushev:1986cu,Kazakov:1987vg,Jones:1986vp,Leigh:1995ep}.}

\bigskip
\noindent
{\bf Proof:}\\
Insert $\beta_{ijk}$ as~given by~the  REs~into the~relationship (\ref{rbeta}).
Since~these chiral~anomalies vanish, we~get
for $\beta_g$ an~homogeneous equation~of the~form
\beq
0=\beta_g(1+O(\hbar)).
\label{prooftheo}
\eeq
The~solution of~this equation~in the~sense of~a formal~power series~in
$\hbar$ is $\beta_g=0$, order~by order.  Therefore, due~to the~REs (\ref{redeq2}), $\beta_{ijk}=0$ too.

Thus~we see~that finiteness~and reduction~of couplings~are intimately~related. Since~an equation~like \refeq{rbeta} is~absent in~non-SUSY theories, one~cannot extend~the  validity~of a~similar theorem~in such~theories.

A~very interesting~development was~done in \cite{Kazakov:1997nf}.
Based~on the~all-loop relations~among the $\beta$-functions of~the soft~SUSY breaking~terms
and~those of~the rigid~supersymmetric theory~with the~help of~the differential~operators,
discussed~in \refse{sec:RCN=1}, it~was shown~that certain~RGI surfaces~can be~chosen, so~as to~reach all-loop~finiteness of~the full~theory. More~specifically, it~was shown~that on~certain RGI~surfaces the~partial differential~operators appearing~in \refeq{betaM},(\ref{betah}) acting~on the
$\beta$- and $\gamma$-functions of~the rigid~theory can~be transformed~to total~derivatives.
Then~the all-loop~finiteness of~the $\beta$~and $\gamma$-functions of~the rigid~theory can~be
transferred~to the~$\beta$-functions~of the SSB terms. Therefore, a~totally all-loop~finite $N=1$ SUSY~gauge theory~can be~constructed,
including~the soft~SUSY breaking~terms.


\section{Phenomenologically Interesting Models with Reduced Couplings}

In this section we review the basic properties of phenomenologically~viable SUSY~models that use the idea of reduction of couplings. Their predictions for quark masses, the light Higgs boson mass, the SUSY breaking scale (defined as the geometric mean of stops), $M_S$, the full SUSY spectrum and the Cold Dark Matter (CDM) relic density (in the case the lightest neutralino is considered a CDM candidate) are discussed in Sections \ref{se:minimal}-\ref{se:rmssm}. The set of experimental constraints employed can be found in Section \ref{se:constraints}. Note that in the examination of the various models we use the unified gaugino mass $M$ instead of $M_S$, as a more indicative parameter of scale.

\subsection{The Minimal $N=1$ SUSY $SU(5)$ }\label{sec:minimalsu5}

First, we present the partial~reduction of couplings~in the~minimal $N = 1$ SUSY
 model~based on~the $SU(5)$ \cite{Kubo:1994bj,Kubo:1996js}.
$\Psi^{I}({\bf 10})$ and $\Phi^{I}(\overline{\bf 5})$  accommodate the~three
generations~of quarks~and leptons, $I$ running over~the three~generations, an adjoint $\Sigma({\bf 24})$ breaks $SU(5)$~down to the MSSM gauge group $SU(3)_{\rm C}
\times SU(2)_{\rm L} \times U(1)_{\rm Y}$,  and~$H({\bf 5})$ and~$\overline{H}({\overline{\bf 5}})$ describe the
two Higgs superfields of the electroweak~symmetry~breaking (ESB)  \cite{dimop,sakai}.
Only one set of $({\bf 5} + {\bf \bar{5}})$ is used to describe the Higgs~superfields appropriate for ESB.
This~minimality renders the present version asymptotically~free (negative~$\beta_g$).
Its superpotential is  \cite{dimop,sakai}
\be
\begin{split}
W &= \frac{g_{t}}{4}\,
\epsilon^{\alpha\beta\gamma\delta\tau}\,
\Psi^{(3)}_{\alpha\beta}\Psi^{(3)}_{\gamma\delta}H_{\tau}+
\sqrt{2}g_b\,\Phi^{(3) \alpha}
\Psi^{(3)}_{\alpha\beta}\overline{H}^{\beta}+
\frac{g_{\lambda}}{3}\,\Sigma_{\alpha}^{\beta}
\Sigma_{\beta}^{\gamma}\Sigma_{\gamma}^{\alpha}+
g_{f}\,\overline{H}^{\alpha}\Sigma_{\alpha}^{\beta} H_{\beta}\\
&+ \frac{\mu_{\Sigma}}{2}\,
\Sigma_{\alpha}^{\gamma}\Sigma_{\gamma}^{\alpha}+
\mu_{H}\,\overline{H}^{\alpha} H_{\alpha}~.
\end{split}
\ee
 where~ $t,b$ and $f$ are indices of the antisymmetric ${\bf 10}$ and adjoint ${\bf 24}$ tensors, $\alpha,\beta,\ldots$ are $SU(5)$~indices, and the first two generations Yukawa couplings have been suppressed.
The SSB Lagrangian is
\be
\begin{split}
-{\cal L}_{\rm soft} &=
m_{H_u}^{2}{\hat H}^{* \alpha}{\hat H}_{\alpha}
+m_{H_d}^{2}
\hat{\overline {H}}^{*}_{\alpha}\hat{\overline {H}}^{\alpha}
+m_{\Sigma}^{2}{\hat \Sigma}^{\dag~\alpha}_{\beta}
{\hat \Sigma}_{\alpha}^{\beta}
+\sum_{I=1,2,3}\,[\,
m_{\Phi^I}^{2}{\hat \Phi}^{* ~(I)}_{\alpha}{\hat \Phi}^{(I)\alpha}\\
& +\,m_{\Psi^I}^{2}{\hat \Psi}^{\dag~(I)\alpha\beta}
{\hat \Psi}^{(I)}_{\beta\alpha}\,]
+\{ \,
 \frac{1}{2}M\lambda \lambda+
B_H\hat{\overline {H}}^{\alpha}{\hat H}_{\alpha}
+B_{\Sigma}{\hat \Sigma}^{\alpha}_{\beta}
{\hat \Sigma}_{\alpha}^{\beta}
+h_{f}\,\hat{\overline{H}}^{\alpha}
{\hat \Sigma}_{\alpha}^{\beta} {\hat H}_{\beta}\\
& +\frac{h_{\lambda}}{3}\,{\hat \Sigma}_{\alpha}^{\beta}
{\hat \Sigma}_{\beta}^{\gamma}{\hat \Sigma}_{\gamma}^{\alpha}+
\frac{h_{t}}{4}\,
\epsilon^{\alpha\beta\gamma\delta\tau}\,
{\hat \Psi}^{(3)}_{\alpha\beta}
{\hat \Psi}^{(3)}_{\gamma\delta}{\hat H}_{\tau}+
\sqrt{2}h_{b}\,{\hat \Phi}^{(3) \alpha}
{\hat \Psi^{(3)}}_{\alpha\beta}\hat{\overline{H}}^{\beta}
+\mbox{h.c.}\, \}~,
\end{split}
\ee
where the hat denotes the scalar
components of the chiral~superfields.
The $\beta$- and $\gamma$-functions and a detailed presentation of the model can be found in  \cite{mondragon1} and in \cite {polonsky1,Kazakov:1995cy}.

The minimal~number of SSB terms that do not violate perturbative renormalizability is required in the reduced~theory.
The  perturbatively~unified SSB parameters
significantly~differ from the~universal ones.
The gauge~coupling $g$ is assumed to be the primary~coupling. We should note that the dimensionless sector admits reduction~solutions that are independent of the dimensionful~sector.
Two sets of asymptotically~free (AF) solutions can achieve
a Gauge-Yukawa Unification in~this model \cite{mondragon1}:
\be
\label{two_sol}
\begin{split}
a & : g_t=\sqrt{\frac{2533}{2605}} g + \order{g^3}~,~
g_b=\sqrt{\frac{1491}{2605}} g + \order{g^3}~,~
g_{\lambda}=0~,~
g_f=\sqrt{\frac{560}{521}} g + \order{g^3}~,\\
b & : g_t=\sqrt{\frac{89}{65}} g + \order{g^3}~,~
g_b=\sqrt{\frac{63}{65}} g + \order{g^3}~,~
g_{\lambda}=0~,~g_f=0~.
\end{split}
\ee
The higher~order terms~denote uniquely~computable
power~series in $g$.
These solutions
describe~the boundaries~of an
AF RGI surface in the parameter space, on~which $g_{\lambda}$ and $g_f$
may differ from~zero. This fact makes possible a partial~reduction
where $g_{\lambda}$ and $g_f$ are (non-vanishing) independent~parameters without endangering
AF. The proton-decay safe region of that surface favours solution~$a$.
Therefore,
we choose to be exactly at~the boundary~defined by solution $a$
\footnote{
$ g_{\lambda}=0 $ is inconsistent, but $g_{\lambda} < \sim 0.005$
is necessary in order for the proton
decay~constraint \cite{Kubo:1995cg} to be satisfied.
A small $g_{\lambda} $ is expected not to affect the prediction of unification of  SSB~parameters.}.

The reduction
of dimensionful couplings is performed as in \refeq{reduction}.
It is understood that $\mu_{\Sigma}$, $\mu_H$ and 
$M$ cannot~be reduced in a desired~form and they are treated~as independent~parameters.
The lowest-order
reduction~solution is found to be:
\begin{equation}\label{red_sol_1}
B_H = \frac{1029}{521}\,\mu_H M~,~
B_{\Sigma}=-\frac{3100}{521}\,\mu_{\Sigma} M~,
\end{equation}
\be
\begin{split}
\label{red_sol}
h_t &=-g_t\,M~,~h_b =-g_b\,M~,
~h_f =-g_f\,M~,~h_{\lambda}=0~,\\
m_{H_u}^{2} &=-\frac{569}{521} M^{2}~,~
m_{H_d}^{2} =-\frac{460}{521} M^{2}~,
~m_{\Sigma}^{2} = \frac{1550}{521} M^{2}~,\\
m_{\Phi^3}^{2} & = \frac{436}{521} M^{2}~,~
m_{\Phi^{1,2}}^{2} =\frac{8}{5} M^{2}~,~
m_{\Psi^3}^{2} =\frac{545}{521} M^{2}~,~
m_{\Psi^{1,2}}^{2} =\frac{12}{5} M^{2}~.
\end{split}
\ee
The gaugino mass $M$ characterize~the scale of~the SUSY~breaking.
It is noted that we may include
$B_H$ and $B_{\Sigma}$ as independent~parameters
without~changing the one-loop
reduction~solution (\ref{red_sol}). Also note that, although we have found specific relations among the soft scalar masses and the unified gaugino mass, the sum rule still holds.

\subsection{The Finite $N=1$ SUSY $SU(5)$ }\label{sec:finitesu5}

Next, we review an $SU(5)$ gauge theory which is finite (FUT) to all orders, with reduction of couplings applied to the third fermionic generation. This~FUT was selected in the past due to agreement with experimental constraints at the time~\cite{Heinemeyer:2007tz} and predicted the  light Higgs mass between 121--126 GeV almost five~years prior to the~discovery.\footnote{Improved Higgs mass calculations would yield a different interval, still compatible with current experimental data (see below).}
The particle~content consists of three ($\overline{\bf 5} + \bf{10}$) supermultiplets,
a pair for each generation of~quarks and~leptons, four
($\overline{\bf 5} + {\bf 5}$) and~one ${\bf 24}$ considered~as Higgs~supermultiplets.
When the finite~GUT group is broken, the~theory is no
longer~finite, and we are left~with the~MSSM
\cite{Kapetanakis:1992vx,Kubo:1994bj,Kubo:1994xa,Kubo:1995hm,Kubo:1997fi,Mondragon:1993tw}.

A predictive all-order finite GYU~$SU(5)$ model should~also
have~the following~properties:
\begin{enumerate}
\item
One-loop anomalous~dimensions are~diagonal,
i.e., $\gamma_{i}^{(1)\,j} \propto \delta^{j}_{i} $.
\item The fermions in the irreps
 $\overline{\bf 5}_{i},{\bf 10}_i~(i=1,2,3)$ do
 not couple~to the~adjoint ${\bf 24}$.
\item The two Higgs doublets of the MSSM are mostly made~out of a
pair~of Higgs~quintet and anti-quintet, which~couple to the third~generation.
\end{enumerate}

Reduction of couplings enhances the symmetry, and the superpotential is then given by \cite{Kobayashi:1997qx,Mondragon:2009zz}:
\begin{align}
W &= \sum_{i=1}^{3}\,[~\frac{1}{2}g_{i}^{u}
\,{\bf 10}_i{\bf 10}_i H_{i}+
g_{i}^{d}\,{\bf 10}_i \overline{\bf 5}_{i}\,
\overline{H}_{i}~] +
g_{23}^{u}\,{\bf 10}_2{\bf 10}_3 H_{4} \\
 &+g_{23}^{d}\,{\bf 10}_2 \overline{\bf 5}_{3}\,
\overline{H}_{4}+
g_{32}^{d}\,{\bf 10}_3 \overline{\bf 5}_{2}\,
\overline{H}_{4}+
g_{2}^{f}\,H_{2}\,
{\bf 24}\,\overline{H}_{2}+ g_{3}^{f}\,H_{3}\,
{\bf 24}\,\overline{H}_{3}+
\frac{g^{\lambda}}{3}\,({\bf 24})^3~.\nonumber
\label{w-futb}
\end{align}

A more detailed description of the model and its properties can be found in \cite{Kapetanakis:1992vx,Kubo:1994bj,Mondragon:1993tw}. The non-degenerate and isolated~solutions to
$\gamma^{(1)}_{i}=0$ give:
\bea
&& (g_{1}^{u})^2
=\frac{8}{5}~ g^2~, ~(g_{1}^{d})^2
=\frac{6}{5}~g^2~,~
(g_{2}^{u})^2=(g_{3}^{u})^2=\frac{4}{5}~g^2~,\label{zoup-SOL52}\\
&& (g_{2}^{d})^2 = (g_{3}^{d})^2=\frac{3}{5}~g^2~,~
(g_{23}^{u})^2 =\frac{4}{5}~g^2~,~
(g_{23}^{d})^2=(g_{32}^{d})^2=\frac{3}{5}~g^2~,
\nonumber\\
&& (g^{\lambda})^2 =\frac{15}{7}g^2~,~ (g_{2}^{f})^2
=(g_{3}^{f})^2=\frac{1}{2}~g^2~,~ (g_{1}^{f})^2=0~,~
(g_{4}^{f})^2=0~.\nonumber
\eea
Furthermore, we have the $h=-MC$ relation, while from the sum~rule (see Subsection \ref{sec:RCN=1}) we obtain:
\be
m^{2}_{H_u}+
2 m^{2}_{{\bf 10}} =M^2~,~
m^{2}_{H_d}-2m^{2}_{{\bf 10}}=-\frac{M^2}{3}~,~
m^{2}_{\overline{{\bf 5}}}+
3m^{2}_{{\bf 10}}=\frac{4M^2}{3}~.
\label{sumrB}
\end{equation}
This shows that we have~only two free~parameters
$m_{{\bf 10}}$ and $M$ for the dimensionful~sector.

The GUT symmetry breaks to the MSSM, where we want only two Higgs~doublets. This is
achieved with the introduction of appropriate~mass terms~that allow  a
rotation in the Higgs~sector
\cite{Leon:1985jm,Kapetanakis:1992vx,Mondragon:1993tw,Hamidi:1984gd, Jones:1984qd},
that permits only~one pair of Higgs~doublets (which couple mostly to
the third~family) to remain light~and acquire~vacuum expectation~values.
the usual fine tuning to achieve
doublet-triplet splitting helps the model to avoid~fast proton~decay (but this mechanism has differences compared to the one used in the minimal $SU(5)$ because of the extended~Higgs sector of the finite case).

Thus, below the GUT scale we have the MSSM with the first two generations unrestricted, while the third is given by the finiteness~conditions.

\subsection{Finite $SU(N)^3$ Unification}\label{sec:su33}

One can consider the construction of FUTs
that have a product gauge group. Let us consider an $N=1$ theory with a $SU(N)_1 \times SU(N)_2 \times \cdots \times SU(N)_k$ and  $n_f$ copies (number of~families) 
of the
supermultiplets $(N,N^*,1,\dots,1) + (1,N,N^*,\dots,1) +
\cdots + (N^*,1,1,\dots,N)$.  Then, the one-loop $\beta$-function
coefficient of  the RGE of each $SU(N)$
gauge coupling is
\begin{equation}
b = \left( -\frac{11}{3} + \frac{2}{3} \right) N + n_f \left( \frac{2}{3}
 + \frac{1}{3} \right) \left( \frac{1}{2} \right) 2 N = -3 N + n_f
N\,.
\label{3gen}
\end{equation}
The necessary~condition for finiteness is $b=0$, which occurs only for the choice $n_f = 3$.  Thus, it is natural to consider  three families of~quarks and~leptons.

From a phenomenological point of view, the~choice is the $SU(3)_C
\times SU(3)_L \times
SU(3)_R$ model, which is 
discussed in detail in \citere{Ma:2004mi}. The discussion of the general  well-known~example can be found in~\cite{Derujula:1984gu,Lazarides:1993sn,Lazarides:1993uw,Ma:1986we}. 
 The quarks and the leptons of the model transform as follows:
\begin{equation}
  q = \begin{pmatrix} d & u & h \\ d & u & h \\ d & u & h \end{pmatrix}
\sim (3,3^*,1), ~~~
    q^c = \begin{pmatrix} d^c & d^c & d^c \\ u^c & u^c & u^c \\ h^c & h^c & h^c
\end{pmatrix}
    \sim (3^*,1,3),
\label{2quarks}
\end{equation}
\begin{equation}
\lambda = \begin{pmatrix} N & E^c & \nu \\ E & N^c & e \\ \nu^c & e^c & S
\end{pmatrix}
\sim (1,3,3^*),
\label{3leptons}
\end{equation}
where $h$ are down-type quarks that acquire masses close to $M_{GUT}$. We have to impose a cyclic  $Z_3$ symmetry in order to have equal gauge couplings at the GUT scale, i.e.
\begin{equation}
q \to \lambda \to q^c \to q,
\label{15}
\end{equation}
where $q$ and $q^c$ are given in \refeq{2quarks} and $\lambda$ in
\refeq{3leptons}.  Then the vanishing of the one-loop gauge $\beta$-function, which is the first finiteness condition (\ref{1st}), is~satisfied.
This  leads us to	 the second condition, namely the vanishing of the
anomalous dimensions of all superfields \refeq{2nd}.  Let us write down the superpotential first.  For one
family we have just two trilinear invariants that can
be~used in the~superpotential as~follows:
\begin{equation}
f ~Tr (\lambda q^c q) + \frac{1}{6} f' ~\epsilon_{ijk} \epsilon_{abc}
(\lambda_{ia} \lambda_{jb} \lambda_{kc} + q^c_{ia} q^c_{jb} q^c_{kc} +
q_{ia} q_{jb} q_{kc}),
\label{16}
\end{equation}
where $f$ and $f'$ are~the Yukawa~couplings associated~to each~invariant.
The quark and~leptons obtain~masses when~the scalar~parts of~the
superfields $(\tilde N,\tilde N^c)$ obtain~vacuum expectation~values (vevs),
\begin{equation}
m_d = f \langle \tilde N \rangle, ~~ m_u = f \langle \tilde N^c \rangle, ~~
m_e = f' \langle \tilde N \rangle, ~~ m_\nu = f' \langle \tilde N^c \rangle.
\label{18}
\end{equation}

For three~families, the most general superpotential has 11 $f$
couplings and 10 $f'$ couplings. Since anomalous dimensions of each superfield vanish, 9~conditions are imposed on these couplings:
\begin{equation}
\sum_{j,k} f_{ijk} (f_{ljk})^* + \frac{2}{3} \sum_{j,k} f'_{ijk}
(f'_{ljk})^* = \frac{16}{9} g^2 \delta_{il}\,,
\label{19}
\end{equation}
where
\begin{eqnarray}
&& f_{ijk} = f_{jki} = f_{kij}, \label{20}\\
&& f'_{ijk} = f'_{jki} = f'_{kij} = f'_{ikj} = f'_{kji} = f'_{jik}.
\label{21}
\end{eqnarray}
Quarks~and leptons~receive  masses~when  the~scalar part~of the~superfields $\tilde N_{1,2,3}$ and $\tilde N^c_{1,2,3}$ obtain~vevs:
\begin{eqnarray}
&& ({\cal M}_d)_{ij} = \sum_k f_{kij} \langle \tilde N_k \rangle, ~~~
   ({\cal M}_u)_{ij} = \sum_k f_{kij} \langle \tilde N^c_k \rangle, \label{22} \\
&& ({\cal M}_e)_{ij} = \sum_k f'_{kij} \langle \tilde N_k \rangle, ~~~
   ({\cal M}_\nu)_{ij} = \sum_k f'_{kij} \langle \tilde N^c_k \rangle.
\label{23}
\end{eqnarray}

When the FUT breaks at $M_{\rm GUT}$, we are left with the MSSM
\footnote{\cite{Irges:2011de,Irges:2012ze} and refs therein discuss in detail the spontaneous breaking of $SU(3)^3$.}, where
both Higgs~doublets couple maximally to~the third~generation.
 These doublets are the linear~combinations $\tilde N^c = \sum_i a_i \tilde N^c_i$ and
$\tilde N = \sum_i b_i \tilde N_i$ .  For the choice of the particular combinations we can use the appropriate masses in the~superpotential \cite{Leon:1985jm}, since~they are~not
constrained~by the finiteness~conditions.
The FUT breaking leaves remnants in the form of the boundary~conditions on~the
gauge~and Yukawa~couplings, i.e. \refeq{19}, the~$h=-Mf$
relation and~the soft~scalar mass~sum rule~at $M_{\rm GUT}$. The latter takes the  following form in this model:
\begin{eqnarray}
m^2_{H_u} + m^2_{\tilde t^c} + m^2_{\tilde q} = M^2 =
m^2_{H_d} + m^2_{\tilde b^c} + m^2_{\tilde q}~.
\end{eqnarray}

If the solution of \refeq{19} is both unique and \textit{isolated}, the model is finite in all orders. This leads $f'$  to vanish   and we are left with the relations
\begin{equation}
f^2 = f^2_{111} = f^2_{222} = f^2_{333} = \frac{16}{9} g^2\,.
\label{isosol}
\end{equation}
Since all $f'$ parameters are zero in one-loop level, the lepton masses are zero. 
They cannot appear
radiatively (as one would expect) due to the finiteness
conditions, and remain as a
problem for further study.

If the solution is just unique (but not isolated, i.e. parametric) we can keep non-vanishing $f'$ 
and achieve two-loop finiteness, in which case lepton masses are not fixed to zero. Then we have a slightly different set of conditions that restrict the Yukawa couplings:
\begin{eqnarray}
f^2 = r \left(\frac{16}{9}\right) g^2\,,\quad
f'^2 = (1-r) \left(\frac{8}{3}\right) g^2\,,
\label{fprime}
\end{eqnarray}
where  $r$ is free and parametrizes the~different
solutions~to the~finiteness conditions. It is important to note that we use the sum rule as boundary condition to the soft  scalars.

\subsection{Reduction of Couplings in the MSSM}\label{sec:mssm}

Finally, we present a version of the MSSM with reduced couplings. All work is carried out in the framework~of the~MSSM, but with the assumption of a covering~GUT. The original partial reduction in this model was done and analysed in \cite{Mondragon:2013aea,Mondragon:2017hki} and is once more restricted to the third fermionic generation. The superpotential in given by
\be
\label{supot2}
W = Y_tH_2Qt^c+Y_bH_1Qb^c+Y_\tau H_1L\tau^c+ \mu H_1H_2\, ,
\ee
and the SSB Lagrangian is
\be
\label{SSB_L}
\begin{split}
-\mathcal{L}_{\rm SSB} &= \sum_\phi m^2_\phi\hat{\phi^*}\hat{\phi}+
\left[m^2_3\hat{H_1}\hat{H_2}+\sum_{i=1}^3 \frac 12 M_i\lambda_i\lambda_i +\textrm{h.c}\right]\\
&+\left[h_t\hat{H_2}\hat{Q}\hat{t^c}+h_b\hat{H_1}\hat{Q}\hat{b^c}+h_\tau \hat{H_1}\hat{L}\hat{\tau^c}+\textrm{h.c.}\right] ,
\end{split}
\ee
The Yukawa~$Y_{t,b,\tau}$ and~the trilinear~$h_{t,b,\tau}$ couplings correspond only to the third family.

Starting with~the dimensionless sector we consider the top and bottom Yukawa couplings, which will be expressed in terms of the strong coupling. The other  gauge couplings, as well as the tau~Yukawa coupling are treated as~corrections.
The REs give
\[
\frac{Y^2_i}{4\pi}\equiv \al_i=G_i^2\al_3,\qquad i=t,b,
\]
and, using~the Yukawa RGE,
\[
G_i^2=\frac 13 ,\qquad i=t,b.
\]

Furthermore, the above reduction is dictated by the different running behaviour of the couplings of 
$SU(2)$ and $U(1)$  compared to the strong one
\cite{Kubo:1985up}, as well as the incompatibility of including the
tau~Yukawa, since its $G^2$ coefficient
turns~negative~\cite{{MTZ:14}}. Adding all three couplings as
corrections, one obtains
\be
\label{Gt2_Gb2}
G_t^2=\frac 13+\frac{71}{525}\rho_1+\frac 37 \rho_2 +\frac 1{35}\rho_\tau,\qquad
G_b^2=\frac 13+\frac{29}{525}\rho_1+\frac 37 \rho_2 -\frac 6{35}\rho_\tau
\ee
where
\be
\label{r1_r2_rtau}
\rho_{1,2}=\frac{g_{1,2}^2}{g_3^2}=\frac{\al_{1,2}}{\al_3},\qquad
\rho_\tau=\frac{g_\tau^2}{g_3^2}=\frac{\displaystyle{\frac{Y^2_\tau}{4\pi}}}{\al_3}
\ee

\noindent Corrections~in \refeq{Gt2_Gb2} are calculated at the $M_{GUT}$ and assuming
\[
\frac{d}{dg_3}\left(\frac {Y_{t,b}^2}{g_3^2}\right)=0.
\]

\noindent This assumption practically states  that, even~including these corrections,
at $M_{GUT}$ the ratio of the top (or bottom) coupling over~the strong~coupling is constant,
thus they have negligible scale dependence.
This~requirement sets~the boundary condition~at $M_{GUT}$, given in \refeq{Gt2_Gb2}.\\
At two-loop level, we assume the corrections to be of the form
\[
\al_i=G_i^2\al_3+J_i^2 \al_3^2,\qquad i=t,b~,
\]
where the $J_i$'s are
\[
J_i^2=\frac 1{4\pi}\,\frac{17}{24},\qquad i=t,b
\]
when only top, bottom and strong gauge couplings are active. If we switch on the rest of the above-mentioned couplings as corrections, we have
\[
J_t^2=\frac 1{4\pi}\frac{N_t}{D},\quad
J_b^2=\frac 1{4\pi}\frac{N_b}{5D},
\]
where $D$, $N_t$ and $N_b$ are known quantities given in \cite{Heinemeyer:2017gsv}.

Let us now move to the dimensionful couplings of the SSB sector of the Lagrangian, namely~the trilinear~couplings $h_{t,b,\tau}$ given in \refeq{SSB_L}. Following the same pattern as in the dimensionless case,
 we first reduce $h_{t,b}$,~while $h_\tau$ is
treated~as a~correction.
\[
h_i=c_i Y_i M_3 = c_i G_i M_3 g_3,\qquad i=t,b,
\]
with $M_3$ the gluino~mass.
The use of the $h_t$ and $h_b$ RGEs gives
\[
c_t=c_b=-1,
\]
where~we used~the 1-loop~relation between~the gaugino~mass and~the gauge~couplings RGE
\[
2M_i\frac {dg_i}{dt}=g_i\frac {dM_i}{dt},\qquad i=1,2,3.
\]
Switching on the other  gauge couplings and $h_\tau$
  as corrections, we have
\[
c_t=-\frac{A_A A_{bb} + A_{tb} B_B}{A_{bt} A_{tb} - A_{bb} A_{tt}},\qquad
c_b=-\frac{A_A A_{bt} + A_{tt} B_B}{A_{bt} A_{tb} - A_{bb} A_{tt}}.
\]
Again, $A_{tt}$, $A_{bb}$ and $A_{tb}$ are given in \cite{Heinemeyer:2017gsv}.

Finally, we~turn our attention to the soft scalar masses $m^2_\phi$ of the SSB
Lagrangian. Their reduction (see \refse{sec:dimful}) takes
the~form
\be\label{mM_rel}
m_i^2=c_i M_3^2,\quad i=Q,u,d,H_u,H_d.
\ee
Then, the soft scalar masses RGEs at one loop  reduce to the following  (the corrections from the  tau Yukawa, $h_\tau$ and the two gauge~couplings are included)
\[
\begin{split}
c_Q=&-\frac{c_{Q{\rm Num}}}{D_m},\quad
c_u=-\frac 13\frac{c_{u{\rm Num}}}{D_m},\quad
c_d=-\frac{c_{d{\rm Num}}}{D_m},\\
c_{H_u}=&-\frac 23\frac{c_{Hu{\rm Num}}}{D_m},\quad
c_{H_d}=-\frac{c_{Hd{\rm Num}}}{D_m},
\end{split}
\]
where $D_m$, $c_{Q{\rm Num}}$, $c_{u{\rm Num}}$, $c_{d{\rm Num}}$, $c_{Hu{\rm Num}}$, $c_{Hd{\rm Num}}$  and the complete analysis are again given in \cite{Heinemeyer:2017gsv}.

\noindent For the completely~reduced system,~i.e. $g_3,Y_t,Y_b,h_t,h_b$, the~coefficients of~the soft scalar masses~become
\[
c_Q=c_u=c_d=\frac 23,\quad c_{H_u}=c_{H_d}=-1/3,
\]
obeying~the sum~rules
\[
\frac{m_Q^2+m_u^2+m_{H_u}^2}{M_3^2}=c_Q+c_u+c_{H_u}=1,\qquad
\frac{m_Q^2+m_d^2+m_{H_d}^2}{M_3^2}=c_Q+c_d+c_{H_d}=1.
\]

\smallskip

Concerning the gaugino masses, the Hisano-Shiftman relation (\refeq{M-M0})
is applied to each gaugino mass as a boundary condition at the GUT scale, where the gauge couplings are considered unified. Thus, at one-loop level, each gaugino mass is only dependent on the b-coefficients of the gauge $\beta$-functions and the arbitrary $M_0$:
\be
M_i=b_iM_0~.
\ee
This means that we can make a choice of $M_0$ such that the gluino mass equals the unified gaugino mass, and the other two gaugino masses are equal to the gluino mass times the ratio of the appropriate b-coefficients.

In \refse{se:rmssm} we begin with the selection of the free parameters. This discussion is  intimately connected 
to the fermion masses predictions.


\section{Phenomenological Constraints}\label{se:constraints}

In our phenomenological analysis  we apply several experimental constraints, which we will briefly review in this section.

Starting from the quark masses, we calculate the top quark pole mass, while the bottom quark mass is evaluated at $M_Z$, in order not to encounter uncertainties inherent to its pole~mass. Their experimental values are \cite{Tanabashi:2018oca},
\beq
\mb(M_Z) = 2.83 \pm 0.10 \gev ~.
\label{mbexp}
\eeq
and
\beq
\mt^{\rm exp} = (173.1 \pm 0.9) \gev~.
\label{mtexp}
\eeq

The~discovery of~a Higgs-like~particle at~ATLAS and~CMS in~July 2012
\cite{Aad:2012tfa,Chatrchyan:2012xdj} can~be interpreted~as the~discovery of~the light~$\cal
CP$-even Higgs~boson of~the MSSM~Higgs spectrum \cite{Mh125,hifi,hifi2}. The
experimental~average for~the (SM) Higgs~boson mass is
\cite{Tanabashi:2018oca}%
\footnote{This is the latest available LHC combination. More recent
  measurements confirm this value.}
\beq
M_H^{\rm exp}=125.10\pm 0.14~{\rm GeV}~.\label{higgsexpval}
\eeq
The theoretical~accuracy \cite{Degrassi:2002fi,BHHW,Bahl:2019hmm},
however, for the 
prediction~of $\Mh$ in the~MSSM, dominates the uncertainty. 
In our following analysis of each of the models described, we use the new
{\tt FeynHiggs} code~\cite{Degrassi:2002fi,BHHW,FeynHiggs,Bahl:2019hmm}
(Version 2.16.0) 
to predict the Higgs mass. {\tt FeynHiggs} evaluates the Higgs masses using a combination of 
fixed~order diagrammatic~calculations and  resummation~of
the (sub)leading~logarithmic~contributions at~all orders, and thus provides a reliable
evaluation~of $M_h$ even for~large SUSY~scales. The refinements in this
combination (w.r.t. previous versions \cite{FeynHiggs})  result in a
downward~shift of $M_h$ of order ${\cal O}(2~{\rm GeV})$ for
large SUSY masses. This version of {\tt FeynHiggs} computes the uncertainty of the Higgs boson mass point by point. This theoretical uncertainty is added linearly to the experimental error in \refeq{higgsexpval}.

We also consider four types of flavour
constraints,  in which~SUSY has non-negligible impact, namely the
flavour observables $\br(b \to s \ga)$, $\br(B_s \to \mu^+ \mu^-)$, $\br(B_u
\to \tau \nu)$ and $\Delta M_{B_s}$.
Although we do not use the latest experimental values, no major effect would be expected. 
\begin{itemize}
\item
For~the branching~ratio $\br(b \to s \gamma)$ we~take a~value
from~the Heavy~Flavor Averaging~Group (HFAG) \cite{bsgth,HFAG}:
\beq
\frac{\br(b \to s \gamma )^{\rm exp}}{\br(b \to s \gamma )^{\rm SM}} = 1.089 \pm 0.27~.
\label{bsgaexp}
\eeq

\item
For~the branching~ratio $\br(B_s \to \mu^+ \mu^-)$ we~use a~combination of~CMS and~LHCb data \cite{Bobeth:2013uxa,RmmMFV,Aaij:2012nna,CMSBsmm,BsmmComb}:
\beq
\br(B_s \to \mu^+ \mu^-) = (2.9\pm1.4) \times 10^{-9}~.
\eeq

\item
For~the $B_u$ decay~to $\tau\nu$ we~use the~limit~\cite{SuFla,HFAG,PDG14}:
\beq
\frac{\br(B_u\to\tau\nu)^{\rm exp}}{\br(B_u\to\tau\nu)^{\rm SM}}=1.39\pm 0.69~.
\eeq

\item
For~$\Delta M_{B_s}$  we~use \cite{Buras:2000qz,Aaij:2013mpa}:
\beq
\frac{\Delta M_{B_s}^{\rm exp}}{\Delta M_{B_s}^{\rm SM}}=0.97\pm 0.2~.
\eeq
\end{itemize}

We finally consider Cold Dark Matter (CDM) constraints.
Since the lightest neutralino, being the Lightest
SUSY Particle (LSP), is a
very promising candidate for CDM~\cite{EHNOS},
we demand that our LSP is indeed the lightest neutralino  and  we discard parameters~leading to different LSPs.
The current bound on the CDM relic density at
$2\,\sigma$~level is given by~\cite{Komatsu:2010fb,Komatsu:2014ioa}%
\footnote{While this is not the latest value, updates would have no
  visible effect on our analysis.}
\beq
\Omega_{\rm CDM} h^2 = 0.1120 \pm 0.0112~.
\label{cdmexp}
\eeq
For the calculation of the relic density of each model we use the \MO~code ~\cite{Belanger:2001fz,Belanger:2004yn,Barducci:2016pcb}) (Version 5.0). The calculation of annihilation and  coannihilation channels is also included. It should be noted that other CDM constraints do not affect our models significantly, and thus were not included in our analysis.


\section{Numerical Analysis of the Minimal $N=1$ $SU(5)$ }\label{se:minimal}

\begin{figure}[H]
\centering
\includegraphics[width=0.495\textwidth]{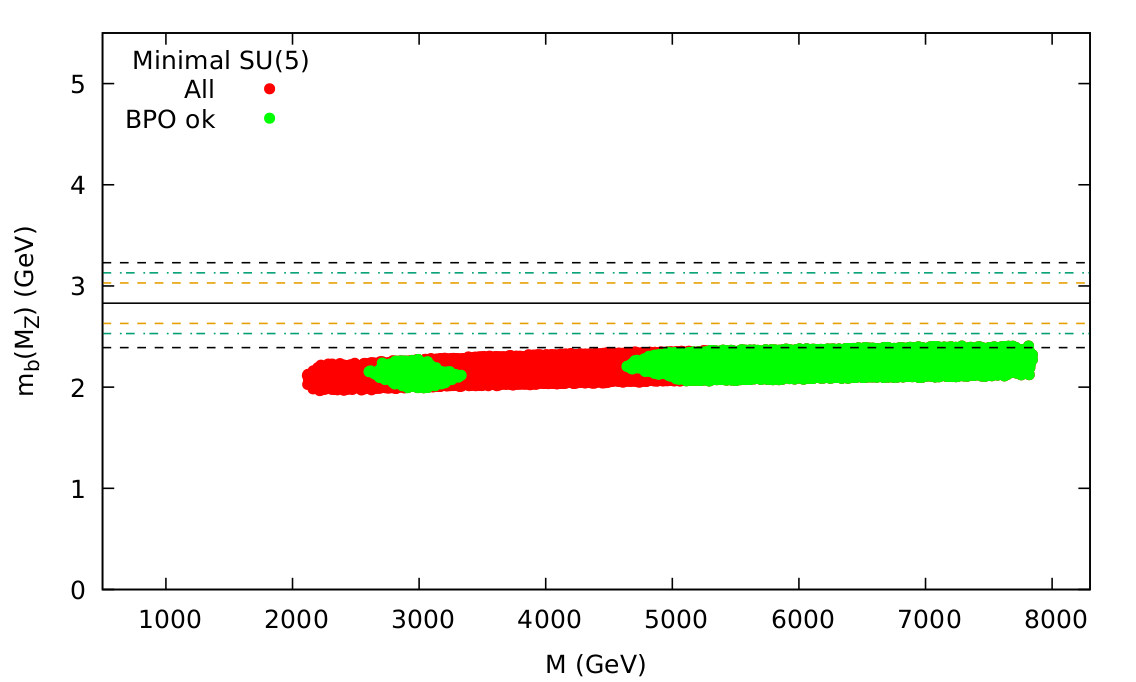}
\includegraphics[width=0.495\textwidth]{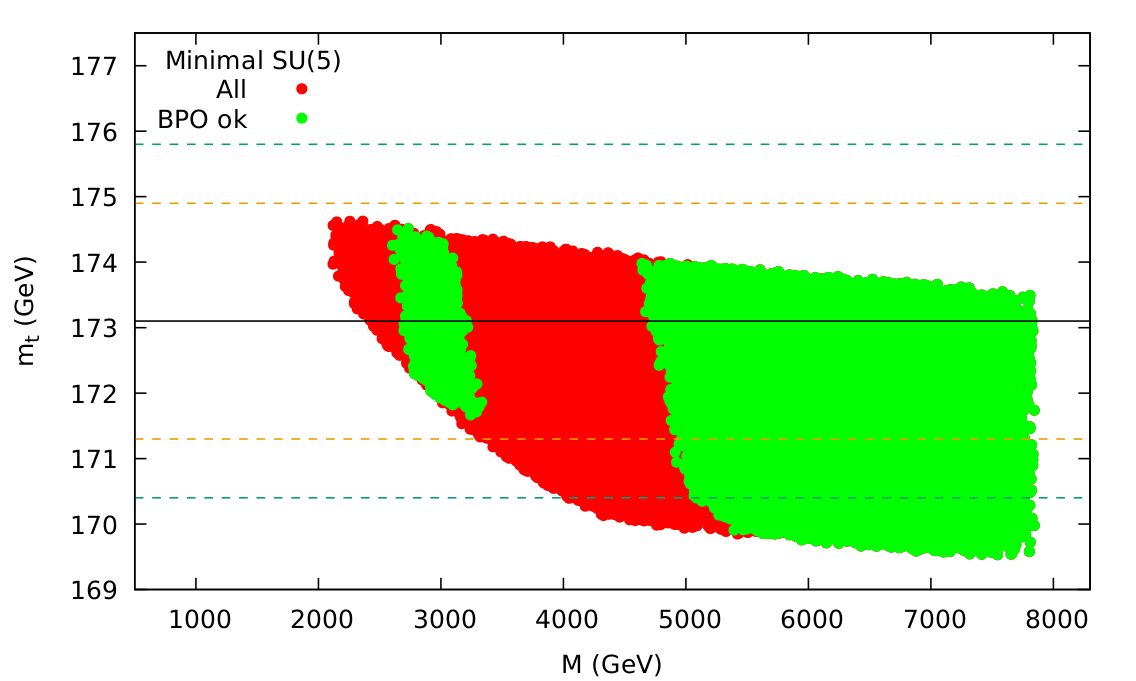}
\caption{\small The bottom quark mass at the $Z$~boson scale (left)
and top quark pole mass (right) are shown
as a function of $M$ for the Minimal $N=1$ $SU(5)$. The green points are
the ones that satisfy the B-physics constraints.
The orange (blue) dashed lines denote the 2$\sigma$ (3$\sigma$) experimental
uncertainties, while the black dashed lines in the left plot add a $\sim
6 \mev$ theory uncertainty to that. \normalsize
}
\label{fig:mintopbotvsM}
\end{figure}

Here, we analyse the particle spectrum predicted by the Minimal $N=1$ SUSY $SU(5)$ as discussed in Subsection \ref{sec:minimalsu5}
for $\mu < 0$.  Below $M_{\rm GUT}$ all couplings and~masses of the theory~run according
to the RGEs of the MSSM.  Thus we examine~the evolution of these
parameters~according to their RGEs up to two-loops for~dimensionless
parameters and at one-loop~for dimensionful ones imposing the
corresponding boundary~conditions. In \reffi{fig:mintopbotvsM}, we
show~the predictions for $\mb (M_Z)$  and $\mt$ as a function of~the
unified gaugino mass $M$. The green points include~the B-physics
constraints. The $\Delta M_{B_s}$ channel is responsible for the gap at
the $B$-physics allowed points.  One can see that, once more, the model
(mostly) prefers the higher energy region of the spectrum (especially
with the admission of $B$-physics constraints).  
The orange (blue) lines denote the 2$\sigma$ (3$\sigma$) experimental
uncertainties, while the black dashed lines in the left plot add a $\sim
6 \mev$ theory 
uncertainty to that. The uncertainty for the boundary conditions of the
Yukawa couplings is taken to be $7\%$, which is included in the spread
of the points shown. 
In the~evaluation of the bottom mass we~have included the
corrections~coming from bottom squark-gluino loops and~top
squark-chargino loops~\cite{Carena:1999py}. 
One can see in the left plot of \reffi{fig:mintopbotvsM} that only by 
taking all uncertainties to their limit, some points at very high~$M$
are within these bounds. I.e.\ confronting the Minimal $N=1$ SUSY
$SU(5)$ with the quark mass measurements ``nearly'' excludes this model,
and only a very heavy spectrum might be in agreement with the
experimental data.

\begin{figure}[H]
\centering
\includegraphics[width=0.495\textwidth]{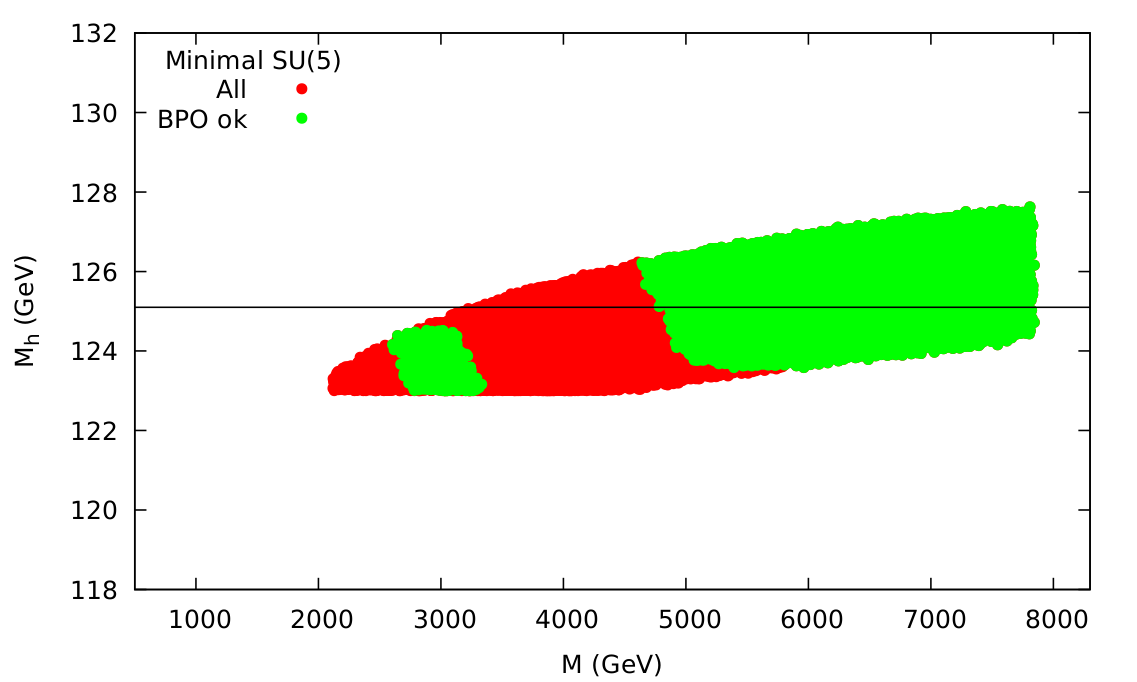}
\includegraphics[width=0.495\textwidth]{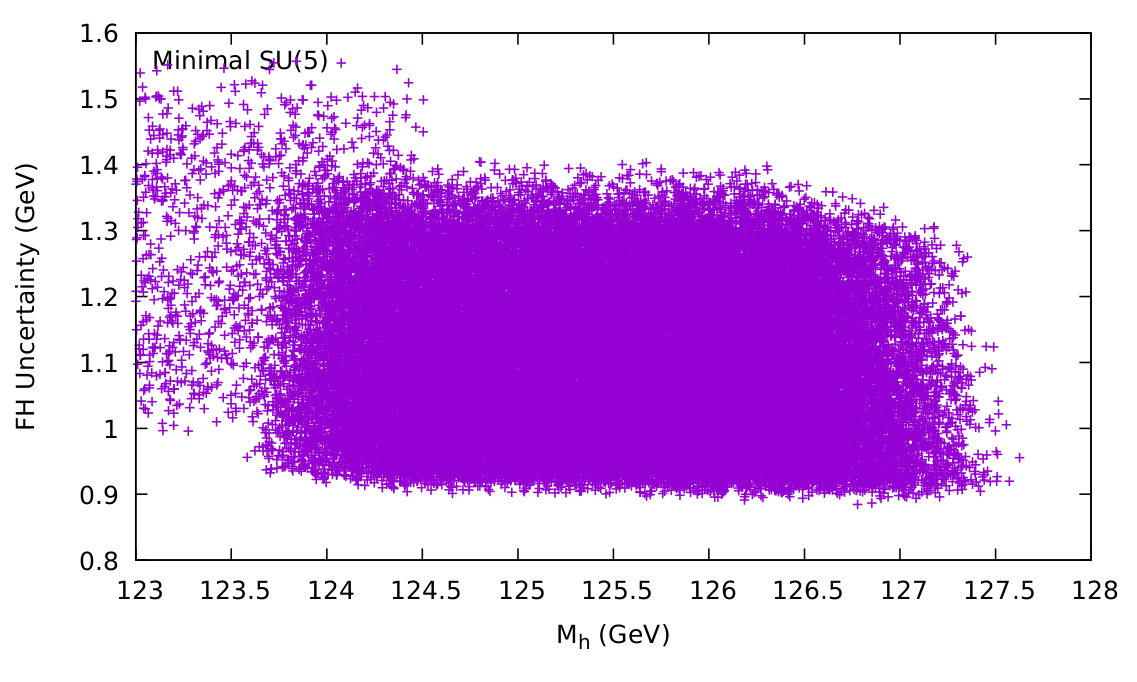}
\caption{\small Left: The lightest Higgs mass, $M_h$, as~a function of $M$ for the Minimal $N=1$ $SU(5)$ model. The $B$-physics constraints allow (mostly) higher scale points (with green colour).  
Right: The lightest Higgs mass theoretical uncertainty \cite{Bahl:2019hmm}. \normalsize}
\label{fig:minhiggsvsM}
\end{figure} 

The prediction for $M_h$ with $\mu < 0$ is given in
\reffi{fig:minhiggsvsM} (left), for a unified gaugino mass between
$2 \tev$ and $8 \tev$, where again the green points satisfy $B$-physics
constraints. \reffi{fig:minhiggsvsM} (right) gives the theoretical
uncertainty of the Higgs mass for each point, calculated with 
{\tt FeynHiggs} 2.16.0~\cite{Bahl:2019hmm}. There is substantial
improvement to the Higgs mass uncertainty compared to past analyses,
since it has dropped by more than $1 \gev$. 

\begin{figure}[htb!]
\centering
\includegraphics[width=0.6\textwidth]{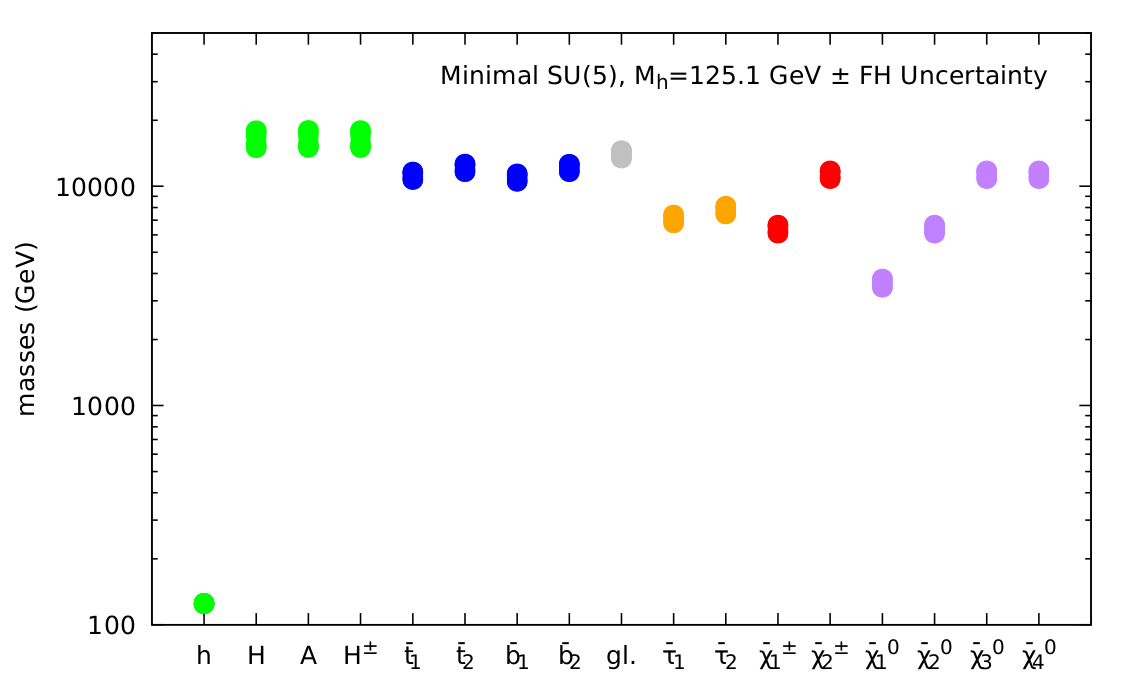}
\caption{\small The plot shows~the spectrum of
    the Minimal $N=1$ $SU(5)$ model for points with Higgs mass within its calculated uncertainty. The green points~are the
    various Higgs boson masses; the blue points are the
    two scalar top and~bottom masses; the gray ones are the gluino
    masses; then come the~scalar tau masses in orange;
    the red points are the~two chargino masses;
    followed~by the purple points indicating the
    neutralino~masses.\normalsize}
\label{fig:minsusyspectrum}
\end{figure}

The full particle spectrum of the model (third generation of fermions
only) that complies with quark mass and B-physics constraints as well as
with the Higgs-boson mass constraint is shown in
\reffi{fig:minsusyspectrum}. Here the points used have a Higgs mass
within the bounds $125.1\pm$~unc, where ``unc'' denotes the uncertainty
shown in \reffi{fig:minhiggsvsM} (right). 
Correspondingly, in \refta{tab:minspectrum} we present an example spectrum,
that is in agreement with all the constraints. The tables shows
the~lightest and the heaviest spectrum (based on $\mneu1$).
The Higgs boson~masses are denoted as $\Mh$, $\MH$, $\MA$ and
$\MHp$. $m_{\tilde{t}_{1,2}}$, $m_{\tilde{t}_{1,2}}$, $\mgl$ and
$m_{\tilde{\tau}_{1,2}}$, are the~scalar top, bottom, gluino~and tau
masses,~respectively. $\mcha{1,2}$ and $\mneu{1,2,3,4}$ stand for
chargino and neutralino~masses, respectively.
As expected from the quark~mass discussion, one can observe that the
allowed spectrum is extremely heavy. Depending on the details, the
FCC-hh might be able to observe some parts of the (colored)
spectrum~\cite{fcc-hh}. On the other hand, improved predictions for the
bottom-quark mass may rule out this model, independent of further
experimental data.
 
\begin{table}[htb!]
\renewcommand{\arraystretch}{1.5}
\centering
\begin{tabular}{|c|rrrrrrrrr|}
\hline
 & $\Mh$ & $\MH$ & $\MA$ & $\MHp$ & $m_{\tilde{t}_1}$ & $m_{\tilde{t}_2}$ &
  $m_{\tilde{b}_1}$& $m_{\tilde{b}_2}$ & $\mgl$ \\
\hline
lightest & 124.6 & 15163 & 15163 & 15163 & 10755 & 11683 & 10551 & 11683 & 13477 \\
heaviest & 125.3 & 17920 & 17920 & 17920 & 11609 & 12609 & 11390 & 12615 & 14532 \\

\hline

\hline
 & $m_{\tilde{\tau}_1}$ & $m_{\tilde{\tau}_2}$ &
  $\mcha1$ & $\mcha2$ & $\mneu1$ & $\mneu2$ & $\mneu3$ & $\mneu4$ & $\tb$ \\
\hline
lightest & 6819 & 7486 & 6125 & 10873 & 3468 & 6126 & 10870 & 10873 & 49 \\
heaviest & 7396 & 8107 & 6645 & 11745 & 3772 & 6647 & 11747 & 111752 & 49.5 \\

\hline
\end{tabular}

\caption{\small
Example spectrum of the Minimal $N=1$ $SU(5)$ .
Masses are in~GeV and rounded to 1 (0.1)~GeV (for the~light Higgs mass).\normalsize}
\label{tab:minspectrum}
\renewcommand{\arraystretch}{1.0}
\end{table}

Furthermore, no point fulfills~the
strict bound of \refeq{cdmexp}, since the relic~abundance turns out to be too~high.
Thus, our model needs a mechanism that can
reduce the CDM~abundance in the early~universe. This~issue could~be related~to the~problem of neutrino~masses.
These~masses cannot~be generated~naturally in this particular model,
although~a non-zero value for~neutrino masses has been
established~\cite{PDG14}. However, the model
could be, in~principle, extended~by introducing~bilinear R-parity
violating~terms and introduce~neutrino masses~\cite{Valle:1998bs,Valle3}.
R-parity~violation \cite{herbi}
would have~a small impact~on the above~collider phenomenology~(apart from the fact~that supersymmmetry~search strategies could~not rely on~a
`missing energy' signature), but remove~the CDM bound~of
\refeq{cdmexp} completely.  Other mechanisms, not
involving R-parity~violation and keeping~the `missing energy' signature, that~could be invoked if the~amount of~CDM appears to~be
too~large, concern~the cosmology of~the early~universe. For~example,
``thermal inflation''~\cite{thermalinf}~or ``late time~entropy
injection'' \cite{latetimeentropy} can~bring the~CDM density~into
agreement~with WMAP~measurements.


\section{Numerical Analysis of the Finite $N=1$ $SU(5)$ }\label{se:futb}

In this section we discuss the full particle spectrum predicted in the
Finite $N=1$  SUSY $SU(5)$ model, as discussed in Subsection
\ref{sec:finitesu5}. The gauge symmetry breaks spontaneously below the
GUT scale, so 
conditions set by finiteness do~not restrict the ~renormalization
properties at low~energies. We are left with boundary conditions
on the~gauge and Yukawa~couplings
(\ref{zoup-SOL52}), the~$h=-MC$ 
relation~and the soft
scalar-mass~sum rule at $M_{\rm GUT}$. Again, the uncertainty for the
boundary conditions of the Yukawa couplings is at $7\%$, which again is
included in the spread of the points. 

\begin{figure}[H]
\centering
\includegraphics[width=0.495\textwidth]{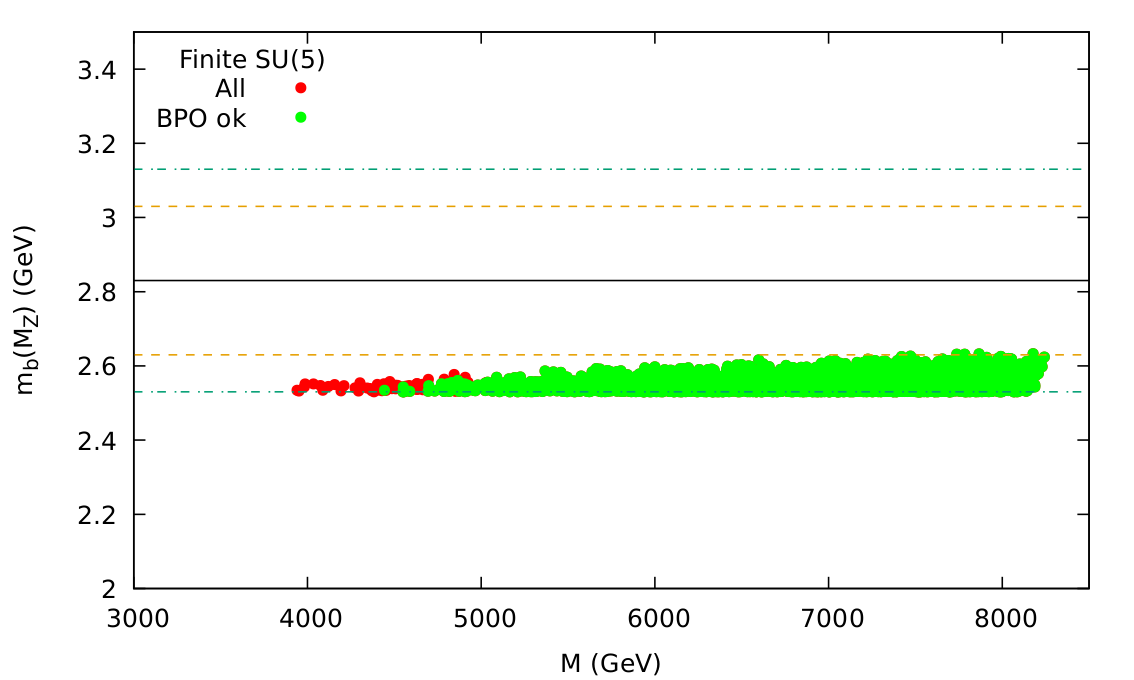}
\includegraphics[width=0.495\textwidth]{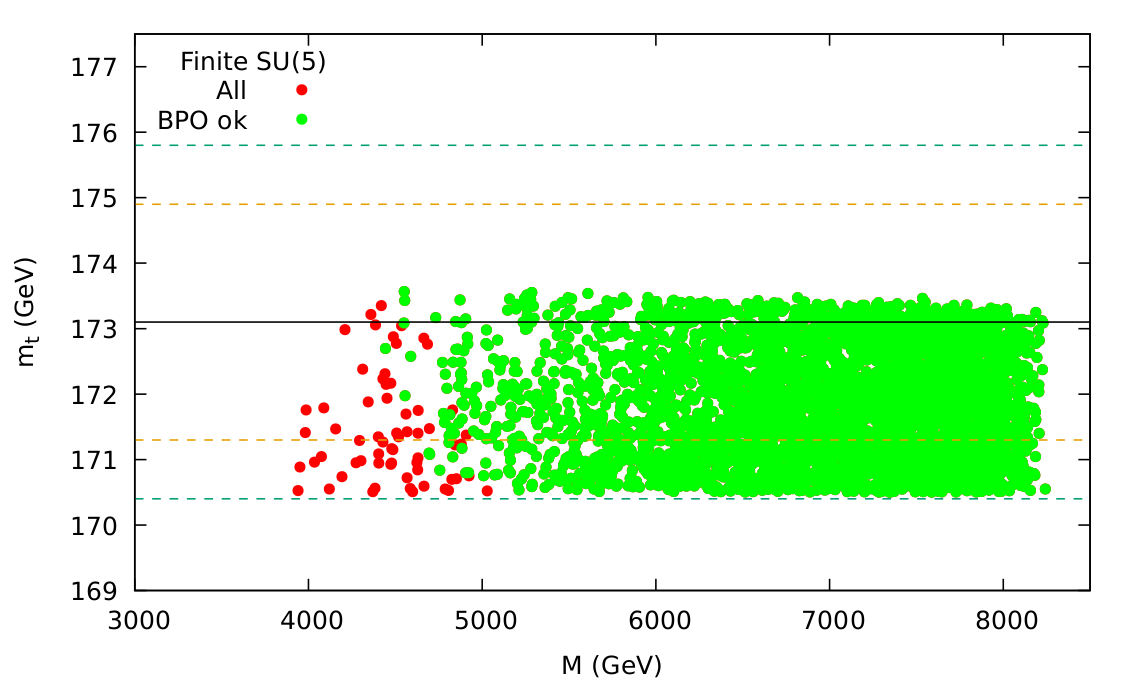}
\caption{\small $\mb (M_Z)$ ({left}) 
and $\mt$ ({right}) as a function of $M$ for the Finite $N=1$ $SU(5)$,
with the color coding as in \protect\reffi{fig:mintopbotvsM}.\normalsize}
\label{fig:futtopbotvsM}
\end{figure}

In \reffi{fig:futtopbotvsM}, $\mb (M_Z)$ and $\mt$ are shown
as~functions of the unified~gaugino mass $M$, where the green points
satisfy the B-physics constraints with the same color coding as in
\reffi{fig:mintopbotvsM}. Here we omitted the additional theoretical
uncertainty of $\sim 6 \mev$. The only phenomenologically viable
option is to consider $\mu < 0$, as is shown in earlier
work~\cite{Heinemeyer:2012yj,Heinemeyer:2012ai,Heinemeyer:2013fga}. The 
experimental~values are indicated by the~horizontal lines with
the~uncertainties at the $2\,\sig$ and  $3\,\sig$ level. The value of
the bottom mass is lower than in past analyses, sending the allowed energy
scale higher. Also the top-quark mass turns out slightly lower than in
previous analyses. 

\begin{figure}[htb!]
\centering
\includegraphics[width=0.495\textwidth]{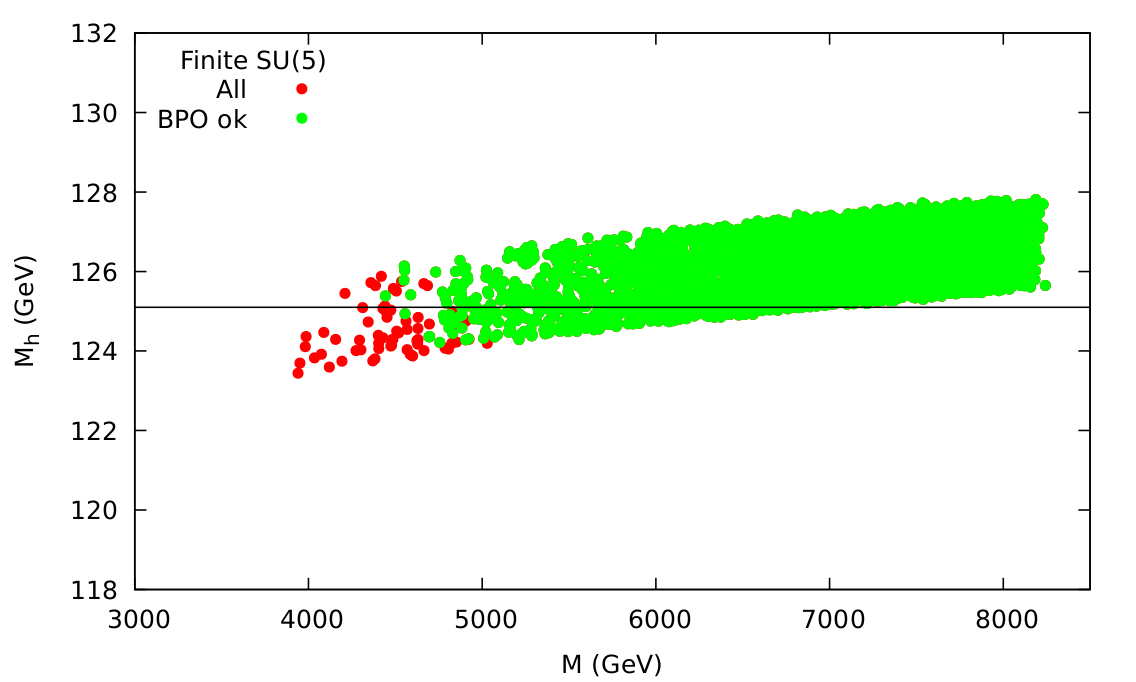}
\includegraphics[width=0.495\textwidth]{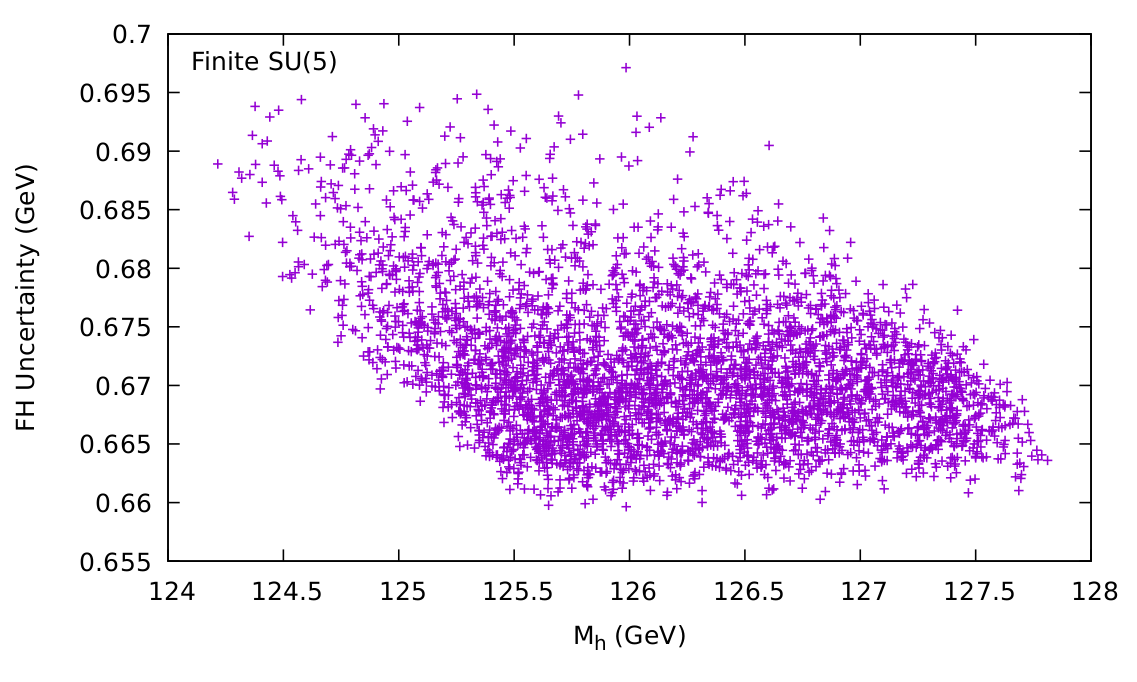}
\caption{\small Left: $M_h$ as~a
    function of $M$. Green points comply with $B$-physics~constraints. Right: The lightest Higgs mass theoretical uncertainty calculated with {\tt FeynHiggs} 2.16.0 \cite{Bahl:2019hmm}.\normalsize}
\label{fig:futhiggsvsM}
\end{figure}

The light Higgs~boson mass is given in \reffi{fig:futhiggsvsM} (left) as
a function~of the unified gaugino mass. Like in the previous section,
these predictions are subject to a theory~uncertainty
\cite{Bahl:2019hmm} that is given in \reffi{fig:futhiggsvsM}
(right). This point-by-point uncertainty (calculated with {\tt
  FeynHiggs}) drops significantly from the flat estimate of 2 and 3~GeV
of past analyses to the much improved $0.65-0.70 \gev$. The $B$-physics
constraints (green points) and the smaller Higgs uncertainty drive the
energy scale above $\sim4.5 \tev$. 
Older analyses,~including in particular less~refined evaluations of the~light
Higgs mass, are given in
\citeres{Heinemeyer:2012yj,Heinemeyer:2012ai,Heinemeyer:2013fga}.
It should be noted that, w.r.t.\ previous analyses the top-quark mass
turns out to be slightly lower. Consequently, higher scalar top masses
have to be reached in order to yield the Higgs-boson mass around it's
central value of \refeq{higgsexpval}, resulting in a correspondingly
heavier spectrum. 

\begin{figure}[htb!]
\centering
\includegraphics[width=0.6\textwidth]{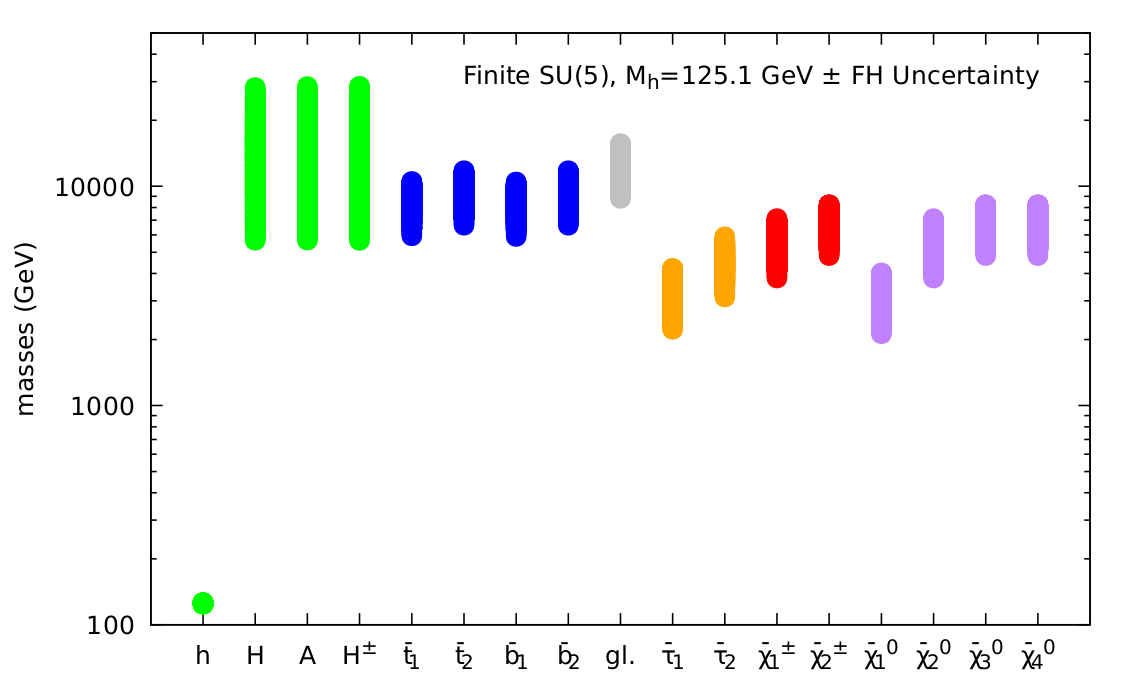}
\caption{\small The plot shows the spectrum of
    the Finite $N=1$ $SU(5)$ model for points in agreement with all
    experimental constraints (see text). The color coding is as in 
\protect\reffi{fig:minsusyspectrum}.\normalsize}
\label{fig:futsusyspectrum}
\end{figure}

In \reffi{fig:futsusyspectrum} we show the full particle spectrum (for
the third fermionic generation), where we only keep points that fulfill
all the experimental constraints (see above). 
Correspondingly, in \refta{tab:futspectrum} we give an example spectrum
which shows the mass range of the parameter space that complies with all
the above-mentioned experimental constraints.
Compared to our previous~analyses 
\cite{Heinemeyer:2010xt,Heinemeyer:2012yj,Heinemeyer:2013nza,Heinemeyer:2012ai,Heinemeyer:2013fga,Heinemeyer:2018roq},
the improved evaluation of $\Mh$ and its uncertainty, together with a
lower prediction of the top-quark mass prefers a heavie Higgs and SUSY spectrum.
In particular, very heavy coloured SUSY particles are favoured (nearly
independent of the $\Mh$ uncertainty), in agreement with
the non-observation of those~particles at the LHC~\cite{2018:59}.
Overall, the~allowed coloured SUSY masses would remain
unobservable~at the HL-LHC, the~ILC or CLIC. However,
the coloured spectrum~would be accessible at the FCC-hh \cite{fcc-hh}, as could
the lower part of the heavy Higgs-boson spectrum.

\begin{table}[htb!]
\renewcommand{\arraystretch}{1.5}
\centering
\begin{tabular}{|c|rrrrrrrrr|}
\hline
 & $\Mh$ & $\MH$ & $\MA$ & $\MHp$ & $m_{\tilde{t}_1}$ & $m_{\tilde{t}_2}$ &
  $m_{\tilde{b}_1}$& $m_{\tilde{b}_2}$ & $\mgl$ \\
\hline
lightest & 124.4 & 5513 & 5513 & 5510 & 5940 & 6617 & 5888 & 6617 & 8819 \\
heaviest & 125.8 & 28121 & 28121 & 28120 & 10486 & 11699 & 10318 & 11686 & 15509 \\

\hline

\hline
 & $m_{\tilde{\tau}_1}$ & $m_{\tilde{\tau}_2}$ &
  $\mcha1$ & $\mcha2$ & $\mneu1$ & $\mneu2$ & $\mneu3$ & $\mneu4$ & $\tb$ \\
\hline
lightest & 2225 & 3123 & 3819 & 4801 & 2120 & 3811 & 4820 & 4811 & 50 \\
heaviest & 4215 & 5788 & 7108 & 8200 & 4019 & 7108 & 8227 & 8227 & 51 \\

\hline
\end{tabular}

\caption{\small
Example spectrum of the Finite $N=1$ $SU(5)$ .
Masses are in~GeV and rounded to 1 (0.1)~GeV (for the~light Higgs mass).\normalsize}
\label{tab:futspectrum}
\renewcommand{\arraystretch}{1.0}
\end{table}

\newpage

Concerning DM, the model exhibits a high relic abundance for CDM. The
CDM alternatives proposed for the Minimal $SU(5)$ model can also be
applied here. It should be noted that the bilinear R-parity violating
terms proposed in the previous section preserve finiteness, as well.


\section{Numerical Analysis of the Two-Loop Finite $N=1$ $SU(3)\otimes SU(3)\otimes SU(3)$ }\label{se:trinification}

We continue our analysis with the two-loop finite $N=1$ SUSY $SU(3)\otimes SU(3)\otimes SU(3)$ model, as described  in the Subsection \ref{sec:su33}. Again, below $M_{\rm GUT}$ we get the MSSM. We further assume a~unique
SUSY breaking scale $M_{\rm SUSY}$ and~below that scale the
effective theory~is just the SM. The boundary condition uncertainty is at $5\%$ for the Yukawa couplings and at $1\%$ for the strong gauge coupling and the soft parameters.

We take into account two new thresholds for the masses of the new particles $h$'s and $E$'s (of the third family in particular) at $\sim10^{13} \gev$ and $\sim 10^{14} \gev$. This results in a wider phenomenologically viable parameter space~\cite{Mondragon:2011zzb}.
Specifically, one of the down-like exotic particles decouples at
$10^{14} \gev$, while the rest  decouple at $10^{13} \gev$.

\begin{figure}[htb!]
\begin{center}
\includegraphics[width=0.495\textwidth]{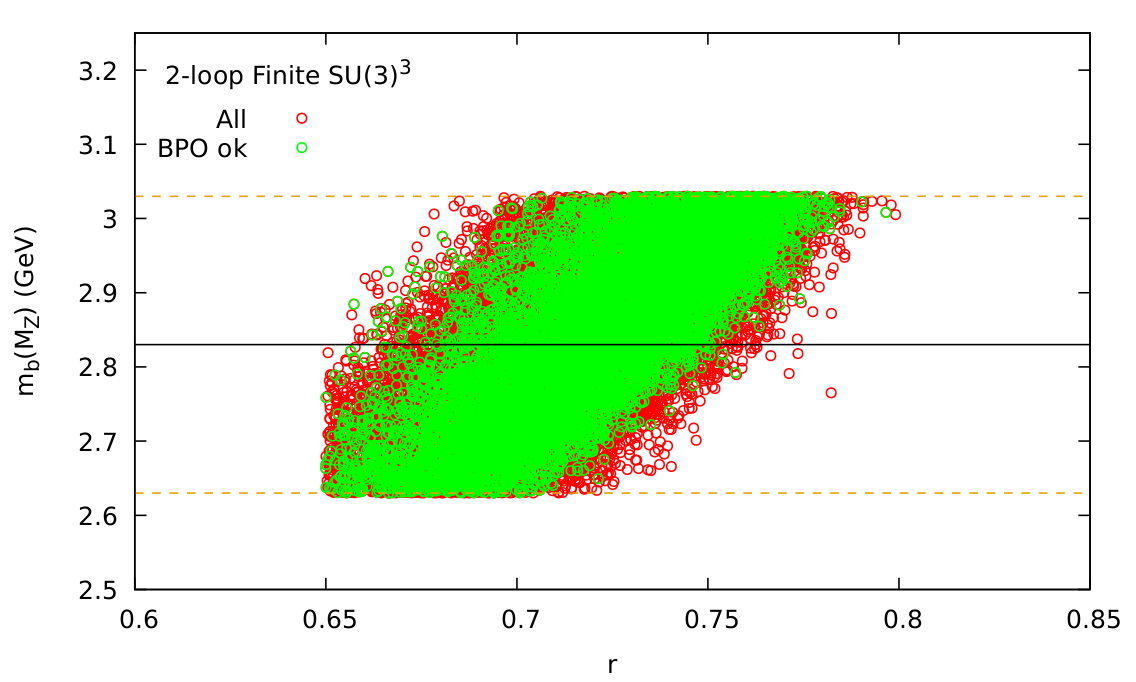}
\includegraphics[width=0.495\textwidth]{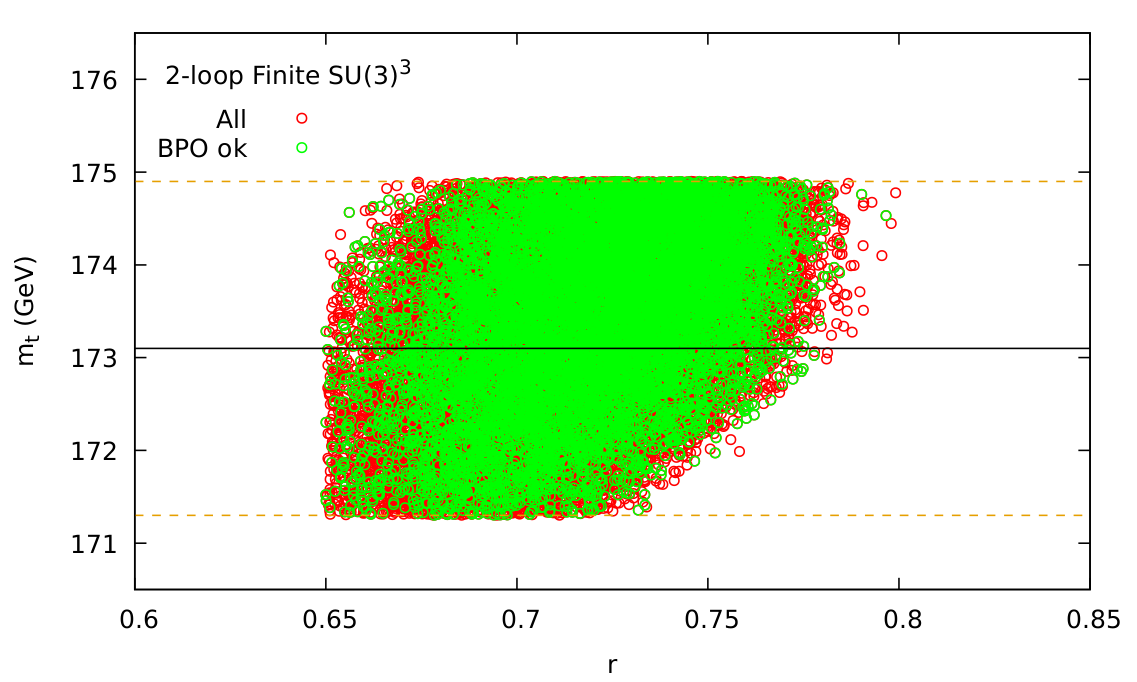}
\caption{\small Bottom and top quark
  masses for the Finite $N=1$ $SU(3)\otimes SU(3)\otimes SU(3)$ model, with $\mu <0 $, as functions of
  $r$. The color coding is as in \protect\reffi{fig:mintopbotvsM}.\normalsize}
\label{fig:su33tbr}
\end{center}
\end{figure}

We compare our predictions with the experimental value of $m_t^{\rm exp}$, while in the case of the bottom quark we take again the value evaluated at
$M_Z$, see \refeq{mbexp}. We single out the $\mu < 0$ case as the most promising model.
With the inclusion of thresholds for the decoupling of the exotic particles, the parameter space allowed predicts a top quark mass in agreement with experimental bounds (see \refeq{mtexp}), which is an important improvement from past versions of the model \cite{Ma:2004mi,Heinemeyer:2009zs,Heinemeyer:2010zzb,Heinemeyer:2010zza}. 
Looking for the values of the~parameter
$r$ (see  Subsection \ref{sec:su33}) which comply with the~experimental
limits (see Section \ref{se:constraints}) for $m_b(M_Z)$  and $m_t$, we
find, as shown in \reffi{fig:su33tbr}, that both masses are in~the experimental range for the same value of $r$ between $0.65$ and $0.80$.  It is important to note that the two~masses are simultaneously~within two sigmas of the experimental bounds.

\begin{figure}[htb!]
\centering
\includegraphics[width=0.495\textwidth]{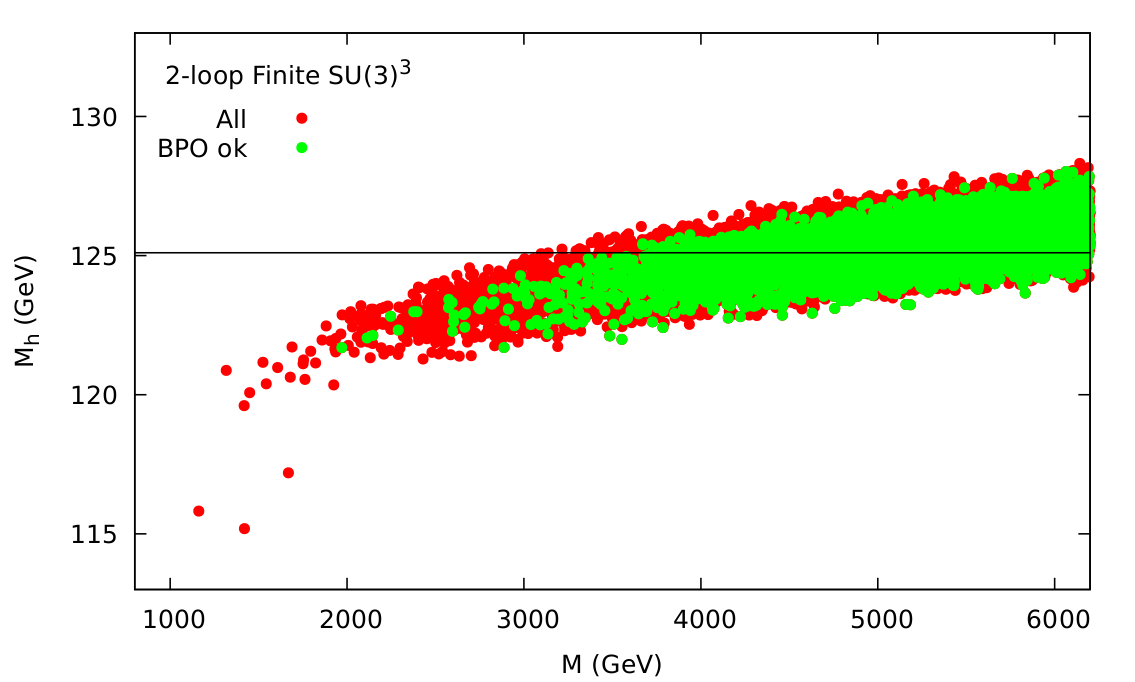}
\includegraphics[width=0.495\textwidth]{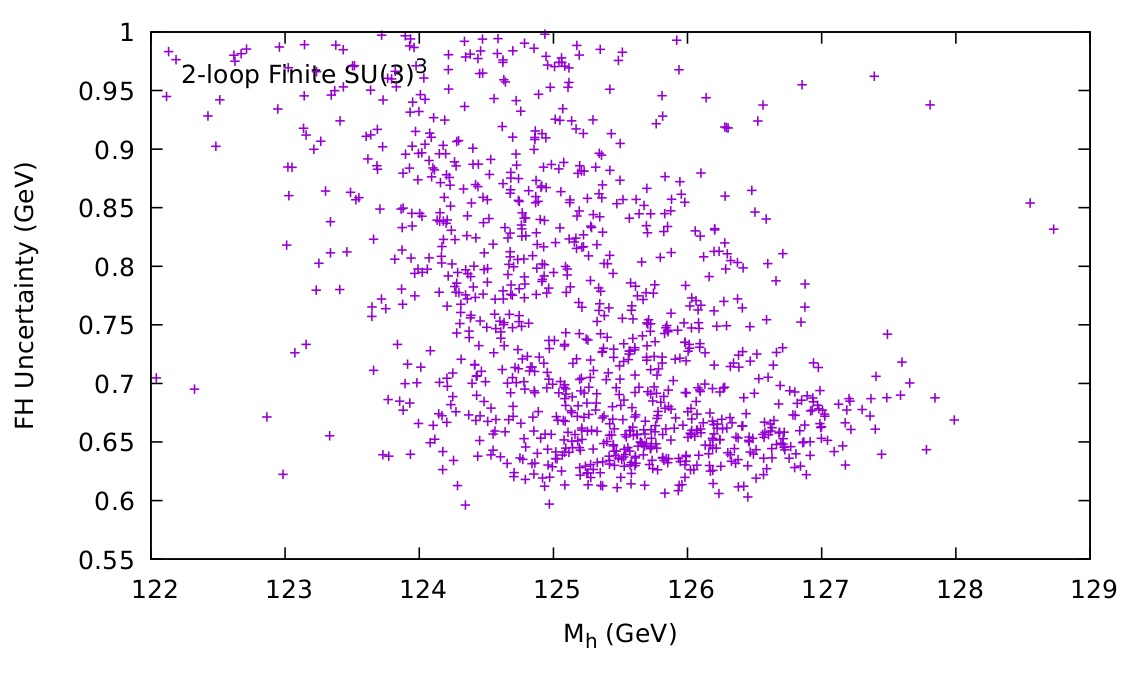}
\caption{\small Left: $M_h$ as~a
    function of $M$ for the Finite $N=1$ $SU(3)\otimes SU(3)\otimes SU(3)$. Right: The Higgs mass theoretical uncertainty \cite{Bahl:2019hmm}.\normalsize}
\label{fig:su33higgsvsM}
\end{figure}

\begin{figure}[htb!]
\centering
\includegraphics[width=0.6\textwidth]{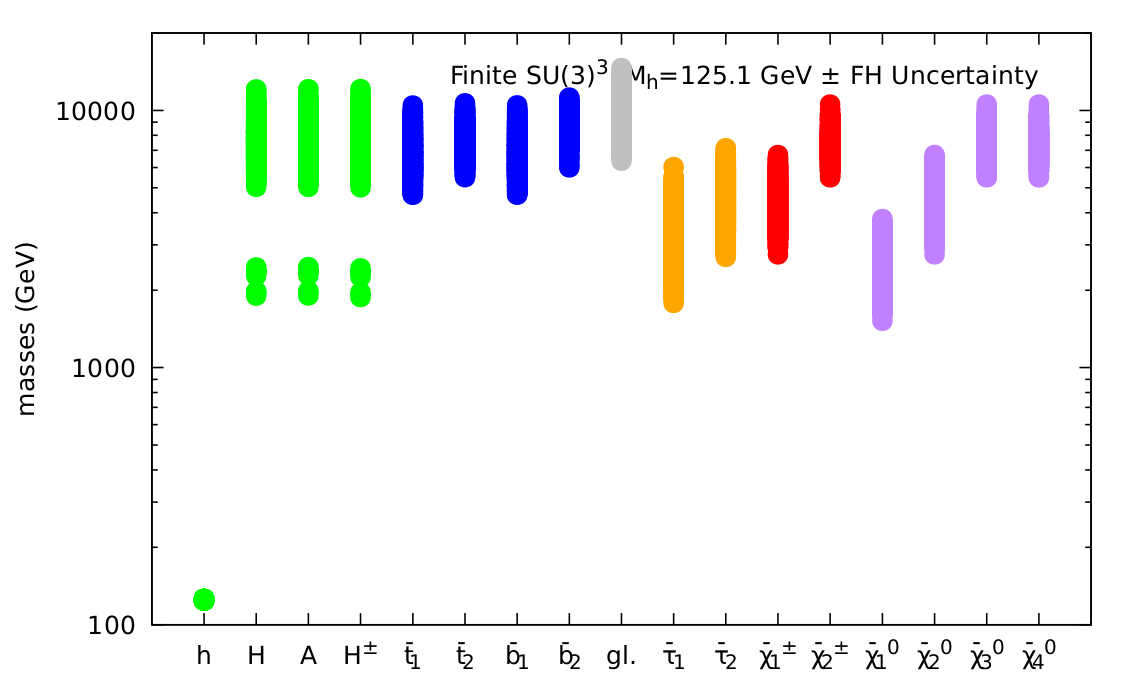}
\caption{\small The spectrum of
    the Finite $N=1$ $SU(3)\otimes SU(3)\otimes SU(3)$ model for points
    with light Higgs mass that satisfies its calculated theoretical
    uncertainty. The color coding is as in
    \protect\reffi{fig:minsusyspectrum}.\normalsize} 
\label{fig:su33susyspectrum}
\end{figure}

In \reffi{fig:su33higgsvsM} (left) the light Higgs boson mass is shown
as a function of the unified gaugino mass, while with the point-by-point
calculated theoretical uncertainty drops below 1~GeV \cite{Bahl:2019hmm}
(\reffi{fig:su33higgsvsM} (right)).  As in the previous models examined,
the $B$-physics constraints (green points in \reffi{fig:su33higgsvsM}
(left) satisfy them) and the new, more restrictive Higgs mass
uncertainty exclude most of the low range of $M$, pushing the particle
spectrum to higher values. This is obvious in \reffi{fig:su33susyspectrum} where the full SUSY spectrum is shown. 
As before, an example spectrum of \refta{tab:su33spectrum} gives the
lightest and heaviest values for each value of the spectrum. 
In fact, all constraints regarding quark masses, the light Higgs boson
mass and B-physics are satisfied, rendering the model very
successful. The only observable that fails to comply with the
experimental bounds is the CDM relic density (see \refeq{cdmexp}). The
lightest neutralino is the LSP and considered as a CDM candidate, but
its relic density does not go below $0.15$, since it is strongly
Bino-like and would require a lower scale of the particle spectrum. It
should be noted that if the B-physics constraints allowed for a unified
gaugino mass $\sim 0.5 \tev$ lower, then agreement with the CDM bounds
as well could be achieved.

\begin{table}[htb!]
\renewcommand{\arraystretch}{1.5}
\centering
\begin{tabular}{|c|rrrrrrrrr|}
\hline
 & $\Mh$ & $\MH$ & $\MA$ & $\MHp$ & $m_{\tilde{t}_1}$ & $m_{\tilde{t}_2}$ &
  $m_{\tilde{b}_1}$& $m_{\tilde{b}_2}$ & $\mgl$ \\
\hline
lightest & 124.2 & 1918 & 1918 & 1917 & 4703 & 5480 & 4671 & 6013 & 6329 \\
heaviest & 125.9 & 12053 & 12053 & 12050 & 10426 & 10631 & 10426 & 11193 & 14550 \\

\hline

\hline
 & $m_{\tilde{\tau}_1}$ & $m_{\tilde{\tau}_2}$ &
  $\mcha1$ & $\mcha2$ & $\mneu1$ & $\mneu2$ & $\mneu3$ & $\mneu4$ & $\tb$ \\
\hline
lightest & 1774 & 2694 & 2736 & 5469 & 1517 & 2736 & 5480 & 5481 & 44 \\
heaviest & 5999 & 7113 & 6713 & 10522 & 3767 & 6703 & 10522 & 10523 & 53 \\

\hline
\end{tabular}

\caption{\small
Example spectrum of the Finite $N=1$ $SU(3)\otimes SU(3)\otimes SU(3)$ .
Masses are in GeV and rounded to 1 (0.1)~GeV (for the light Higgs mass).\normalsize}
\label{tab:su33spectrum}
\renewcommand{\arraystretch}{1.0}
\end{table}

The SUSY and Higgs spectrum corresponding to the experimentally allowed
points turns out to be too heavy for current or most future
experiments. The FCC-hh will be able to test most of the spectrum, in
particular for colored particles. However, also here the highest parts
of the allowed parameter space might be inaccessible even to this
collider.


\section{Numerical Analysis of the Reduced MSSM}\label{se:rmssm}

The relations among reduced~parameters in terms
of the~fundamental ones derived in  \refse{sec:mssm}
have an RGI part and a part that originates from the corrections, and thus  scale dependent. In the present
analysis we choose~the unification scale to apply the~corrections
to all these RGI~relations.
As was noted earlier, the Hisano-Shiftman relation sets a hierarchy among the gaugino masses, rendering Wino the lightest of them. As such, we have a Wino-like lightest neutralino (which is the LSP).

In the dimensionless sector of~the
theory, since $Y_\tau$ is not reduced in favour of the fundamental
parameter $\al_3$, the tau lepton mass is an input parameter and,
consequently, $\rho_\tau$ is an independent~parameter, too.  At low
energies we fix  $\rho_{\tau}$ and $\tan\beta$ using
the~mass of the tau~lepton $m_{\tau}(M_Z)=1.7462$ GeV.
Then, we determine the top and bottom  masses using the value found
for $\tan\beta$ together with $G_{t,b}$, as obtained~from the
REs and their corrections. 

Correspondingly, concerning the dimensionful sector,  $h_\tau$ cannot be expressed in terms of the unified gaugino~mass scale, leaving  $\rho_{h_\tau}$ a free parameter.
  $\mu$ is a free~parameter as well, as it cannot~be reduced in favour of $M_3$
as discussed~above. On the other hand, $m_3^2$ could be~reduced, but here
we choose to leave~it free.
However, $\mu$ and $m_3^2$ are~restricted from the requirement of
EWSB, and only $\mu$ is taken~as an independent~parameter.
Finally, the~other parameter in the Higgs-boson sector,~the $\cp$-odd
Higgs-boson mass $\MA$ is evaluated~from $\mu$, as well as from $m_{H_u}^2$~and
$m_{H_d}^2$, which~are obtained from the REs.
In total, we vary the~parameters $\rho_\tau$, $\rho_{h_\tau}$, $M$
  and~$\mu$.
  
  \begin{figure}[htb!]
\begin{center}
\includegraphics[width=0.495\textwidth]{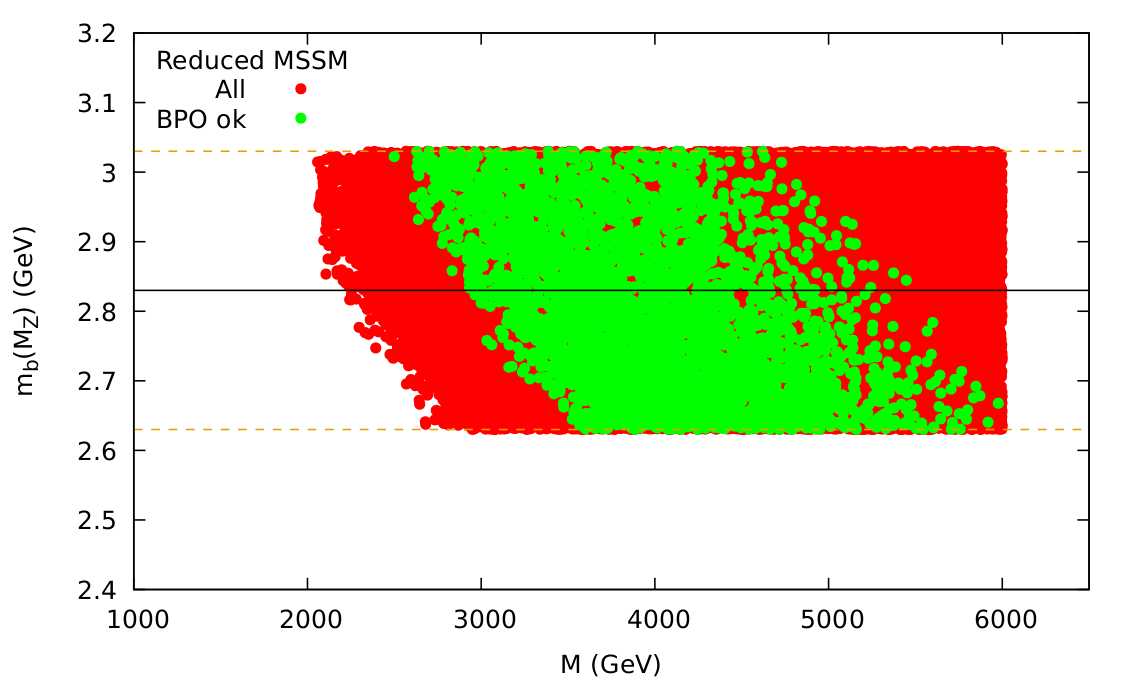}
\includegraphics[width=0.495\textwidth]{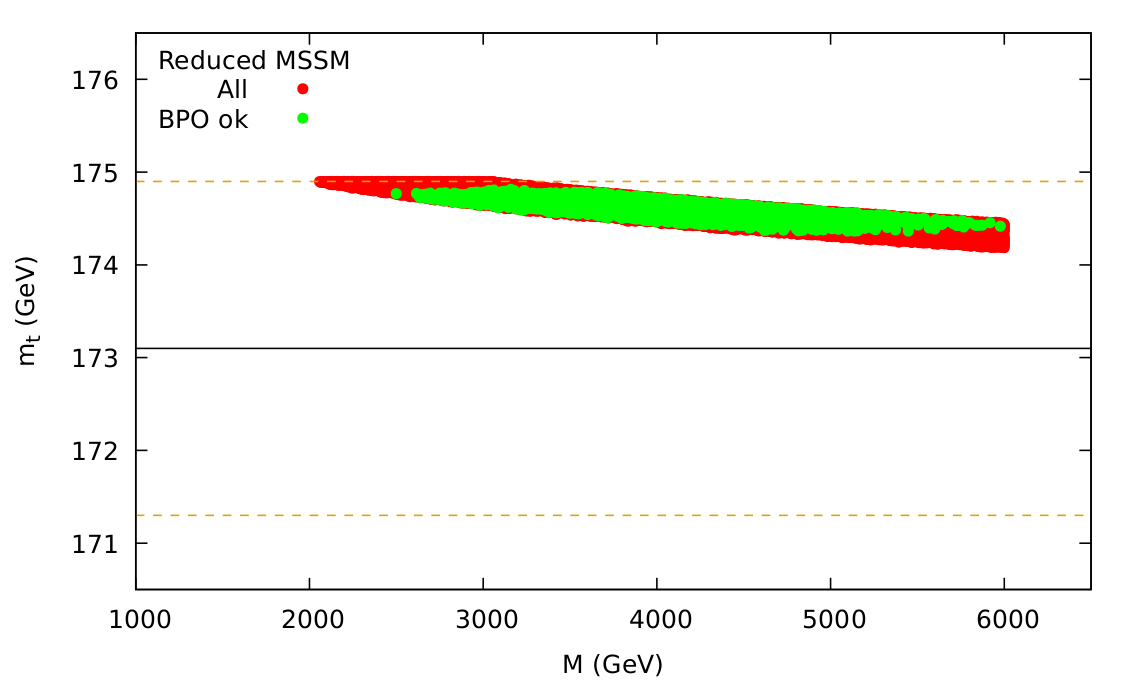}
\caption{\small The left (right) plot shows the bottom (top)~quark mass for the Reduced
    MSSM, with the color coding as in \protect\reffi{fig:mintopbotvsM}.\normalsize
}
\label{fig:rmssmtopbotvsM}
\end{center}
\vspace{-1em}
\end{figure}

As we have already mentioned, the variation~of $\rho_\tau$ gives the running
bottom~quark mass at the $Z$~boson mass scale and the top~pole mass,~where points not within $2\sigma$ of
the experimental~data are neglected, as it is~shown in
\reffi{fig:rmssmtopbotvsM}. The experimental~values (see \refse{se:constraints})
are denoted by the
horizontal lines~with the uncertainties at the $2\,\sig$ level. 
The green dots satisfy the flavour constraints. One can see that the
scan yields many parameter points that
are in very good~agreement with the experimental data and give restrictions in the allowed range of~$M$ (the common gaugino mass at the unification scale).

\begin{figure}[htb!]
\begin{center}
\includegraphics[width=0.495\textwidth]{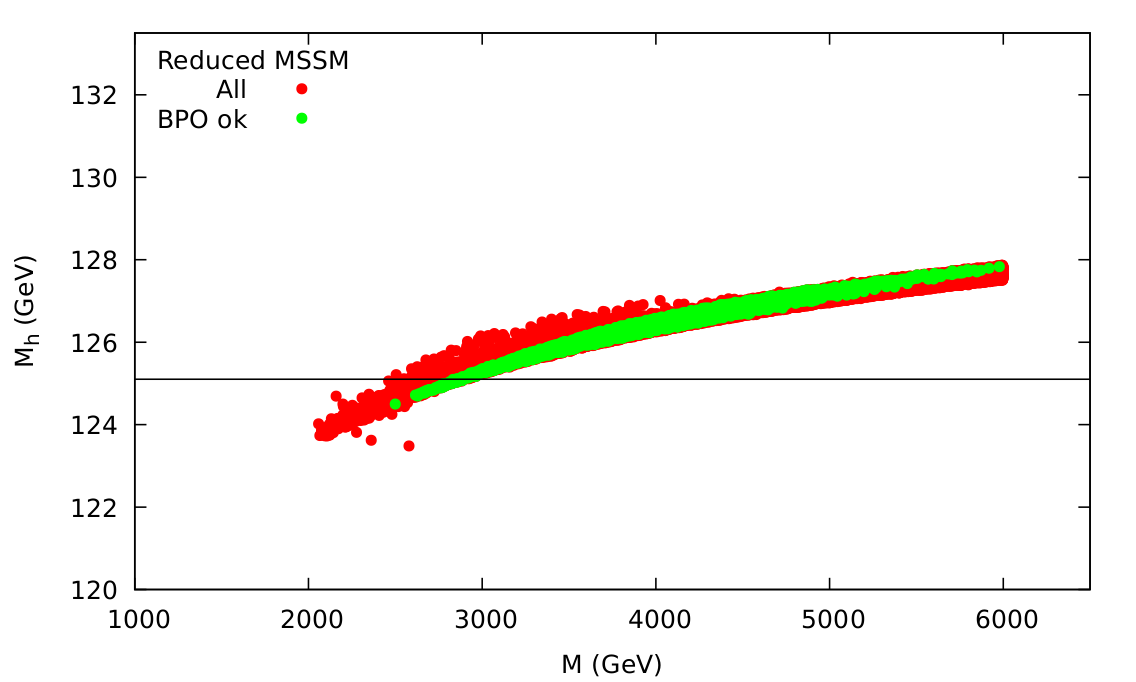}
\includegraphics[width=0.495\textwidth]{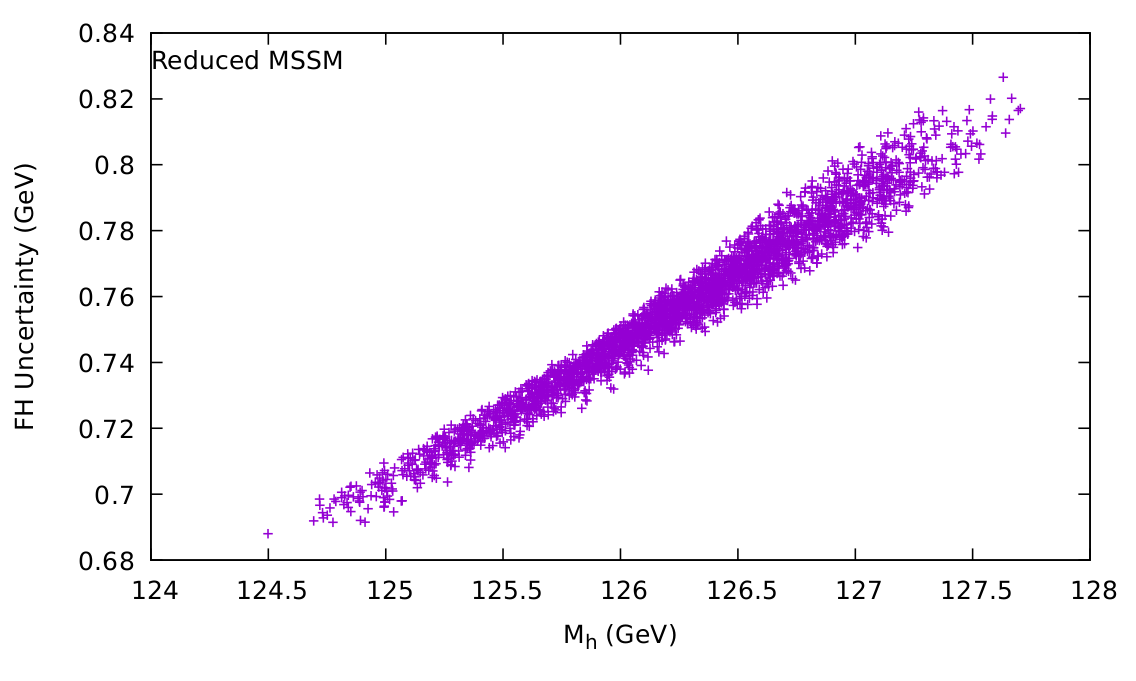}
\caption{\small Left: The lightest Higgs boson mass, $\Mh$ in the Reduced MSSM. The green points is the full model prediction. Right: the lightest Higgs mass theoretical uncertainty \cite{Bahl:2019hmm}.\normalsize}
\label{fig:rmssmhiggsvsM}
\end{center}
\end{figure}   
   
\begin{figure}[htb!]
\begin{center}
\includegraphics[width=0.55\textwidth]{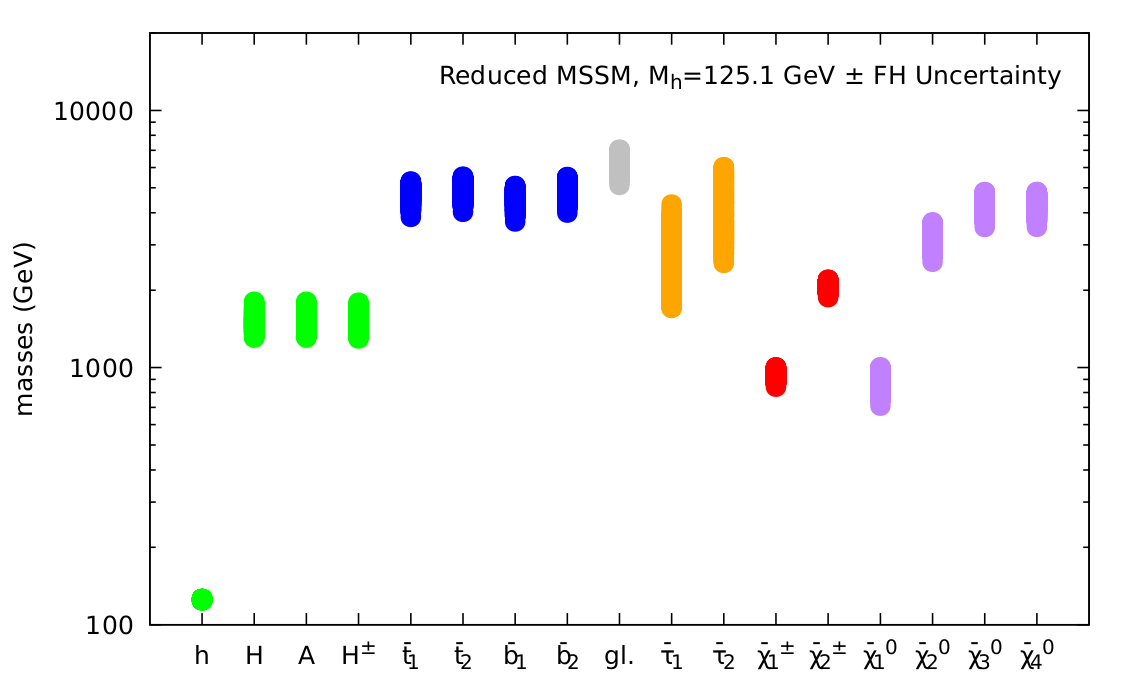}
\caption{\small The full spectrum of the Reduced MSSM
after keeping only the points with Higgs mass that complies with its theoretical uncertainty.
The colours denoting each mass are as described in \refse{se:minimal}.\normalsize
}
\label{fig:rmssmsusyspectrum}
\end{center}
\end{figure}

The prediction for $\Mh$ is shown in
\reffi{fig:rmssmhiggsvsM} (left).
Once again, one should~keep in mind that the theory
uncertainty given in \reffi{fig:rmssmhiggsvsM} (right) has dropped below 1~GeV \cite{Bahl:2019hmm}.  The Higgs mass predicted by the model is  
in the range~measured at the LHC, favoring this time relatively small
values of~$M$. This in turn
sets a limit on the low-energy
SUSY masses, rendering the Reduced MSSM  highly~predictive and
testable. In \reffi{fig:rmssmsusyspectrum} we show its full~spectrum 
(again, third generation of sfermions only), which complies with the
$B$-physics and the Higgs mass uncertainty (with the color coding as in
\reffi{fig:minsusyspectrum}). 
Correspondingly, in \refta{tab:rmssmspectrum} we show an example
spectrum of the lightest and heaviest value of each parameter of the
SUSY spectrum of the Reduced MSSM, in agreement with the Higgs-boson
mass measurement and its calculated theoretical uncertainty, as well as
with the $B$-physics constraints.

From the spectra shown in \reffi{fig:rmssmsusyspectrum} and
\refta{tab:rmssmspectrum} it can be concluded that already 
the HL-LHC~\cite{HAtautau-HL-LHC} will be able to test the full Higgs~spectrum. 
The lighter SUSY particles, which are given by the electroweak spectrum, will mostly remain unobservable~at the LHC
and at future $e^+e^-$~colliders such~as the ILC or CLIC. An exception
are the lightest neutralino and chargino masses, which could be covered
by CLIC3TeV. The coloured~mass spectrum will remain~unobservable at
the~(HL-)LHC, but could be accessible at the~FCC-hh~\cite{fcc-hh}, which
could either confirm the SUSY 
spectrum of the Reduced MSSM or rule it out.

Concerning the DM predictions, it
should be noted that the Hisano-Shiftman relation imposes a Wino-like
LSP, which unfortunately lowers the CDM relic density below the
boundaries of \refeq{cdmexp}. This renders this model viable if
\refeq{cdmexp} is applied only as an upper limit and additional sources
of CDM are allowed. This is in contrast to the other three models
discussed previously. 
 
\begin{table}[t!]
\renewcommand{\arraystretch}{1.5}
\centering
\begin{tabular}{|c|rrrrrrrrr|}
\hline
 & $\Mh$ & $\MH$ & $\MA$ & $\MHp$ & $m_{\tilde{t}_1}$ & $m_{\tilde{t}_2}$ &
  $m_{\tilde{b}_1}$& $m_{\tilde{b}_2}$ & $\mgl$ \\
\hline
lightest & 124.5 & 1305 & 1305 & 1297 & 3851 & 4029 & 3699 & 4007 & 5126 \\
heaviest & 125.8 & 1801 & 1801 & 1780 & 5275 & 5564 & 5076 & 5502 & 7017 \\

\hline

\hline
 & $m_{\tilde{\tau}_1}$ & $m_{\tilde{\tau}_2}$ &
  $\mcha1$ & $\mcha2$ & $\mneu1$ & $\mneu2$ & $\mneu3$ & $\mneu4$ & $\tb$ \\
\hline
lightest & 1705 & 2536 & 843 & 1875 & 711 & 2579 & 3516 & 3517 & 40 \\
heaviest & 4288 & 6008 & 1004 & 2195 & 1001 & 3666 & 4814 & 4815 & 45 \\

\hline
\end{tabular}
\caption{\small Example spectrum of the Reduced~MSSM. All masses are in GeV and~rounded to 1 (0.1)~GeV (for the light~Higgs mass).\normalsize
}
\label{tab:rmssmspectrum}
\renewcommand{\arraystretch}{1.0}
\end{table}

\section{Conclusions}

In this review we have briefly discussed the ideas concerning the reduction of couplings of renormalizable theories and the theoretical tools which
have been developed to confront the problem. Updates and new results
were given for four specific  models, in which the reduction of
parameters has been theoretically explored and tested against the
experimental data. Important updates w.r.t.\ previous analyses are the
improved Higgs-boson mass predictions as provided by the latest version
of {\tt FeynHiggs} (version 2.16.0), including in particular the
improved uncertainty evaluation. Furthermore, we have evaluated the CDM
predictions of each model with \MO (version 5.0). 
From a phenomenological point of view, the reduction of couplings method
described in the article provides  selection~rules that single out
realistic~GUTs.  
It is also possible to work with the reduction of couplings method directly in the MSSM. In this case, the number of free~parameters is decreased substantially and the model becomes more predictive~\cite{Mondragon:2013aea,Mondragon:2017hki,Heinemeyer:2018zpw,Heinemeyer:2018roq,Heinemeyer:2017gsv}.

We focused our analysis in four models, namely the Minimal $N=1$
$SU(5)$, the Finite $N=1$ $SU(5)$,  the Two-Loop Finite  $N=1$
$SU(3)\otimes SU(3)\otimes SU(3)$ and the Reduced MSSM, which are
presented in  \refse{se:minimal}-\ref{se:rmssm} respectively and share
similar features. The Minimal $N=1$ $SU(5)$ model predicts the top quark
mass and the light Higgs boson mass in agreement with LHC measurements,
as well as the full SUSY spectrum of the MSSM. 
However, concering the bottom-quark mass predictions, relatively small
values are obtained, and agreement with the experimental data can be
found at the $3\,\sigma$~level only if additionally a $\sim 6 \mev$
theory uncertainty is included, favoring an extremely heavy SUSY spectrum.
The Finite $N=1$ $SU(5)$ model, the Finite $N=1$ $SU(3)\otimes
SU(3)\otimes SU(3)$ model and the Reduced MSSM are in natural
agreement with all LHC measurements and searches. 
Concerning the DM predictions, the three former models have an excess of
CDM w.r.t.\ the experimental measurements, while the latter has a lower
relic density than required by experimental searches. This renders this
model viable if the experimental value is applied only as an upper limit
and additional sources 
of CDM are allowed. This is in contrast to the other three models
discussed previously. 

All models predict
relatively heavy spectra, the heavy parts of which evade detection in
present and near-future colliders, with the exception of the lighter
part of the Reduced MSSM spectrum. The Higgs sector of that model can be
fully tested already at the HL-LHC, and the lighter electroweak spectrum
could be covered by CLIC3TeV. On the other hand, the FCC-hh will have
the capacity to test large parts of the 
predicted~parameter spaces of all four models. From this point of view,
the Reduced MSSM is the model with the best prospect, since it allows
the lightest spectrum out of the four models. On the theoretical~side,
the long-term challenge is in the~development of a framework in~which
the above~successes of the field 
theory~models are combined with~gravity.


\subsection*{Acknowledgements}

\noindent We thank  H. Bahl, G. Belanger, F. Boudjema, C. Delaunay, T. Hahn, W. Hollik, J. Kalinowski, W. Kotlarski, D. L\"ust, S. Pukhov and E. Seiler for helpful discussions. MM thanks the CERN Department of Theoretical Physics for their hospitality. GP thanks LAPTh of Annecy for their hospitality. GZ thanks the ITP of~Heidelberg, MPI Munich, CERN Department of Theoretical Physics, IFT Madrid and MPI-AEI for their~hospitality.

The work~of S.H.~is supported~in part~by the~MEINCOP Spain~under Contract~FPA2016-78022-P, in~part by~the
Spanish~Agencia Estatal de Investigaci\'{o}n (AEI), the~EU Fondo Europeo~de Desarrollo~Regional (FEDER)
through the~project FPA2016-78645-P, in~part by the ``Spanish Red Consolider
MultiDark'' FPA2017-90566-REDC, and in~part by the AEI~through the grant~IFT
Centro de~Excelencia Severo Ochoa~SEV-2016-0597.
The work of~M.M. is partly~supported by~UNAM PAPIIT through~Grant IN111518.
The work~of G.P., N.T. and~G.Z. is partially supported~by the COST~actions CA15108 and~CA16201.
GZ has been~supported within the~Excellence Initiative~funded by the German and State~Governments,
at the Institute for~Theoretical Physics, Heidelberg University~and from the Excellence Grant~Enigmass
of LAPTh.

\end{document}